\documentclass[12pt]{iopart}

\usepackage{graphicx}
\usepackage{iopams}

\def\sign{{\rm sign}}
\def\Rmath{{\bf R}}
\def\barray{\begin{eqnarray}}
\def\earray{\end{eqnarray}}
\def\beq{\begin{equation}}
\def\eeq{\end{equation}}

\begin{document}

\title{General covariant $xp$ models and the Riemann zeros}

\author{Germ\'an Sierra}
\address{Instituto de F\'{\i}sica Te\'orica, UAM-CSIC, Madrid, Spain}

\eads{
\mailto{german.sierra@uam.es}}

\begin{abstract}
We study  a general class of models whose classical Hamiltonians
are given by $H = U(x) p + V(x)/p$, where $x$ and $p$ are the position and momentum
of a particle moving in one dimension, and $U$ and $V$ are positive functions. This class includes the 
Hamiltonians  
 $H_{\rm I} =x (p+1/p)$ and $H_{\rm II}=(x+ 1/x)(p+ 1/p)$, which have been recently  discussed 
 in connection with the non trivial zeros of the Riemann zeta function. 
We show that all  these models  are covariant under general coordinate
transformations. This remarkable property becomes explicit in the Lagrangian
formulation which describes  a relativistic particle moving in a 1+1 dimensional
spacetime whose metric is constructed from the functions $U$ and $V$. 
 General covariance is  maintained by  quantization and we find 
 that the  spectra  are  closely related
to   the geometry of the associated  spacetimes.  In particular, the Hamiltonian  $H_{\rm I}$ 
corresponds to  a flat spacetime,   whereas its  spectrum  approaches   the
Riemann zeros in average. The latter property  also holds for  the model $H_{\rm II}$,
whose underlying  spacetime is asymptotically flat. These results
 suggest  the existence of a Hamiltonian whose underlying spacetime
encodes  the prime numbers,  and whose spectrum provides   the Riemann zeros.
\end{abstract}


\vspace{2pc}


\maketitle

\section{Introduction}
\label{sec:introduction}

In 1999  Berry and Keating  conjectured  that an appropriate quantization of the 
classical Hamiltonian $H = xp$,  of a particle moving on the real line, 
could provide  the long sought spectral realization of  the Riemann zeros  \cite{BK99,BK99b}. 
These authors  were led to this idea by the similarity between  the semiclassical spectrum of a
 regularized version of the $xp$  model and the
average distribution of the Riemann zeros.  
The regularization introduces the   constraints
$|x| \geq  \ell_x$ and $|p| \geq \ell_p$  in position and momentum,  such that the product
of their  minimal values is equal to the Planck  constant  
($\ell_{x} \ell_p = 2 \pi \hbar)$.  This proposal was made in the framework of Quantum Chaos
and spectral statistics \cite{B86,K99,B03}. 
About the same time,
Connes proposed another regularization of $xp$ based on the constraints
$|x| \leq   \Lambda$ and $|p| \leq \Lambda$, where $\Lambda$ is a common cutoff \cite{C99}. 
In the limit where $\Lambda$ is sent to infinity one obtains a continuum spectrum
where the Riemann zeros are absortion  spectral lines, according to Connes. 
This interpretation  underlies
the adelic approach to the Riemann hypothesis. 
These results have motivated several works in the last  years 
on the $xp$ model,  and  related quantum mechanical models,  
for  their possible connection with the Riemann zeros \cite{A99}-\cite{S11b}
(see \cite{SH11} for a review on  physical approaches to the Riemann hypothesis).

Specially relevant to this paper  are the  recent works   \cite{SL11,BK11}, 
which propose two different modifications of the $xp$ Hamiltonian
in order to have bounded classical trajectories and a discrete quantum spectrum.  
In reference   \cite{SL11}, 
 the classical Hamiltonian is  $H_{\rm I} = x ( p + \ell_p^2/p)$, which adds to $xp$ 
 a non standard  term $x \ell_p^2/p$, where $\ell_p$ is a constant. 
 The latter  term implements, in a dynamical way,  the 
 constraint  $|p| \geq \ell_p$, but one still needs  the constraint
 $x \geq \ell_x$.  The classical Hamiltonian $H_{\rm I}$
 can be quantized in terms of a self-adjoint operator whose spectrum  
 agrees asymptotically with the first two  terms of the Riemann-Mangoldt formula 
 that counts the number of Riemann zeros \cite{E74}. The Hamiltonian $H_{\rm I}$
 breaks the symmetry between $x$ and $p$, which is an appealing 
 feature of the $xp$ model. This fact led Berry and Keating to  propose a new 
 Hamiltonian $H_{\rm II} = (x+ \ell_x^2/x)( p + \ell_p^2/p)$,
 which restores the $x-p$ symmetry and implements dynamically both
 constraints on $x$ and $p$, as can be  seen from the appearance
 of the constants $\ell_{x,p}$ in it.

The aim of this paper is to generalize the previous models,
considering Hamiltonians of  the form   $H = U(x) p + V(x)/p$, where $U(x)$ 
and $V(x)$ are positive functions defined on intervals of the real line.
This  class of Hamiltonians have the remarkable property of being general
covariant,  which means that   they maintain  their  form 
under general coordinate transformations, i.e.  diffeomorphisms  $x'= f(x)$. 
These transformations change the functions $U$ and $V$,  according to prescribed  laws, 
but the physical observables, such as energies,  remain unchanged. 
General  covariance is a signature of gauge symmetry,  as it occurs
in General Relativity. Indeed, we shall show that the present models
describe the motion of a relativistic particle moving in a 1+1 dimensional
spacetime whose  a metric can be constructed in terms of the functions
$U$ and $V$. The classical trajectories of the Hamiltonian $H = U(x) p + V(x)/p$,
being the geodesics of that metric. Hence these generalized $xp$ models
acquire  a geometrical interpretation which gives
new insights into their quantum properties,  and in particular their spectrum.
  
The organization of the paper is as follows. In section \ref{sec:classical}
we  introduce the classical models and show their  general covariance. In section \ref{sec:relativity}
we pass from the Hamiltonian to the Lagrangian formulation and  present 
a relativistic spacetime interpretation, which  is illustrated with several
examples.  In section \ref{sec:trajectory}
we  discuss the classical trajectories in  the Hamiltonian and Lagrangian formulations.
In section \ref{sec:semiclassical} we analyze the semiclassical spectrum of the models
introduced in section \ref{sec:relativity}.  We quantize the models in section \ref{sec:quantization}
and show that general covariance is maintained. Finally,  we present our conclusions. We have included
 in  \ref{sec:abel}  the  derivation of the inverse of the semiclassical quantization formula,  and
 in \ref{sec:kondo} the quantization of  the Hamiltonian $H= p + \ell_p^2/p$.

\section{The classical  Hamiltonian}
\label{sec:classical}

Let us consider a general class of  Hamiltonians of the form

\beq
H = U(x)  \, p + \frac{V(x)}{p},  \qquad x \in D, 
\label{1}
\eeq
where $x$ and $p$ are the position and momentum  of a particle
moving in an interval $D$ of the real line, and $U(x)$ and $V(x)$
are positive functions in $D$. We shall be mainly concerned 
with  intervals that are halflines,  $D= (\ell_x, \infty)$, 
and  eventually with segments  i.e. $D= (\ell_x, \tilde{\ell}_x)$, 
The two examples discussed in the introduction correspond to \cite{SL11,BK11}
 
\barray 
H_{\rm I} & = & x  \left( p + \frac{ \ell_p^2}{p} \right),   \hspace{2.5cm}  D=  (\ell_x, \infty)  
 \label{1a} \\
H_{\rm II} & = & \left( x + \frac{ \ell_x^2}{ x} \right) \left( p + \frac{ \ell_p^2}{ p}  \right), \qquad   D =  (0, \infty).   
\label{1b}  
\earray
Berry and Keating also studied the model (\ref{1b}) on the whole real line, but
we shall not consider this case here because the corresponding functions $U$ and $V$ 
are not positive. 
The positivity conditions on $U$ and $V$ are necessary,  in order to have bounded classical trajectories, 
but not sufficient,  as shown by the example $H = p + \ell_p^2/p$ (see section  \ref{sec:relativity} and \ref{sec:kondo}). 
It is convenient to  write $U$ and $V$ as  
 
\beq
U(x) = u^2(x), \qquad V(x) = v^2(x),  
\label{2}
\eeq
where $u(x)$ and $v(x)$ will also be positive  functions. 
 The   Hamiltonians  (\ref{1}) 
change  their   sign under the time reversal transformation, i.e. 

\beq
x \rightarrow  x, \quad p \rightarrow -p \Longrightarrow H \rightarrow - H, 
\label{2b}
\eeq
which implies that  if   $\{ x(t), p(t) \}$ is a classical trajectory  with energy $E$, so is  $\{ x(t), -p(t) \}$  with energy $-E$.
 Upon quantization,
the  spectrum will contain  time conjugate pairs $\{ E_n, - E_n \}$,  for  appropiate boundary 
conditions  related to the self-adjoint extensions of (\ref{1}).  The breaking of the time reversal symmetry
is suggested  by the statistical properties of  the Riemann zeros, 
that are described by  the Gaussian Unitary Ensemble distribution (GUE) \cite{M74,O89}.

The Hamiltonian (\ref{1}) is covariant under general coordinate transformations
of the variable $x$. Indeed, let us consider the  infinitesimal canonical  transformation

\beq
x' = x + \epsilon(x), \qquad p' = (1 - \partial_x \epsilon(x) )  \,  p, \qquad  |\epsilon(x)| << 1,  
\label{3}
\eeq
that   preserves the Poisson bracket

\beq
\left\{ x, p \right\} = 1 \Longrightarrow \left\{ x', p' \right\} = 1 + O(\epsilon^2). 
\label{4}
\eeq
Substituting these  eqs. into (\ref{1}),  one obtains 

\barray 
H(x,p)   & = &    \left[ U(x') - \epsilon(x') \partial_x U(x' ) + (\partial_{x'} \epsilon(x' )) U(x' ) \right]   p' 
\label{5}   \\
& + &   \left[  V(x') - \epsilon(x') \partial_x V(x' ) -  (\partial_{x'} \epsilon(x' )) V(x' ) \right] \frac{1}{p'} + O(\epsilon^2) 
\nonumber  \\
& = & H(x', p') + O(\epsilon^2),   \nonumber 
\earray
which has the same form as (\ref{1}), for  redefined functions

\barray
U' (x' ) & = &    U(x') - \epsilon(x') \partial_x U(x' ) + (\partial_{x'} \epsilon(x' )) U(x' ), 
 \label{6} \\
V' (x' ) & = &   V(x') - \epsilon(x') \partial_x V(x' ) -  (\partial_{x'} \epsilon(x' )) V(x' ). 
\nonumber 
\earray
These equations are the infinitesimal version of the transformation laws of one dimensional 
 tangent and  cotangent  vectors

\beq
U'(x') =  \left( \frac{d x }{ d x'} \right)^{-1}  U(x(x')), \qquad  V'(x') =   \frac{d x }{ d x'}   V(x(x')). 
\label{9}
\eeq
The momentum $p$ transforms also as a cotangent vector (i.e. one form). Another way to state
these transformations laws is by saying that the products $U(x) (dx)^{-1}, V(x) dx, p dx$
and also $u(x) (dx)^{ -1/2}, v(x) (dx)^{1/2}$ are invariant under reparametrizations of
$x$. To preserve the positivity of the new functions $U'$ and $V'$, 
we shall restrict ourselves to diffeomorphisms $x' = f(x)$,  such that $df(x)/dx >0$. These 
diffeormorphisms  form the group denoted as ${\rm Diff}^+$. The interval  $D$ is mapped into the 
new interval  $D' = f(D)$.  All the models related by diffeomorphisms
are equivalent at the classical level. We shall organize them 
 into equivalent classes described by the quotient 

\beq
{\cal M}_{\rm cl} = \{ U, V, D \}/{\rm Diff}^+. 
\label{9b}
\eeq
Each class can be uniquely characterized by a Hamiltonian which has   a particularly simple form, 

\beq
w(x)= U(x) = V(x) \Longrightarrow H = w(x) \left( p + \frac{1}{p} \right).  
 \label{9c} 
\eeq
Any other Hamiltonian can be brought into this form by a convenient reparametrization. 
For example,  the models (\ref{1a}) and  (\ref{1b}) correspond to 

\barray
H_{\rm I} &  \rightarrow & w_{\rm I}(x) = x, \hspace{1.8cm}  D = (h, \infty), \qquad h = \ell_x \ell_p  \label{10} \\
H_{\rm II} &  \rightarrow & w_{\rm II}(x) = x + \frac{h^2}{ x}, \qquad D = (0, \infty), \qquad h = \ell_x \ell_p  \label{10b}
\earray 
where we made the change of variables $x' = \ell_p x$ in both cases.
The function $w(x)$  
is unique,  up to the  shift $ x \rightarrow x + {\rm cte}$. We shall call the canonical
form (\ref{9c}) the {\em symmetric gauge}. Other 
gauges are possible, as for example $U(x) = 1$, which will be briefly discussed at the end
of the next section. 
$w(x)$ is a scalar function that can be computed in any coordinate system as

\beq
w(x) = u(x) \, v(x).
\label{11}
\eeq
Using the transformation laws of $u(x)$ and $v(x)$ one can 
verify that $w'(x') = w(x)$, as claimed above.  
To find the coordinate transformation that brings a model into the symmetric gauge 
consider the equation

 \beq
 \frac{ v'(x')}{u'(x')} dx'  =   \frac{ v(x)}{ u(x)} dx. 
 \label{12}
 \eeq 
 In the symmetric  gauge  $u'(x') = v'(x')$,  so integrating (\ref{12}) yields
 the mapping  $x' = f(x)$, i.e. 
 
 \beq 
 x' - \ell'_x   = f(x)=  \int_{\ell_x}^x dy \, \frac{ v(y)}{u(y)}
 \label{13}
 \eeq
which is invertible, $x = f^{-1}(x')$, since $v(x)/u(x)>0$. The constant $\ell'_x$
is left undetermined by this map,  so it 
can be choosen at will.  $w'(x')$ is obtained  using eq.(\ref{11})
and the inverse of (\ref{13}) as

\beq
w'(x')  = w(x)= u(x) \, v(x) = u(f^{-1}(x')) \, v(f^{-1}(x')).
\label{14}
\eeq
An  application of   equations (\ref{13}) and (\ref{14}),  is to show
that apparently different models may turn out to be equivalent, 
as shown by the following case. Consider
the model,

\beq
H_{\rm III} =    x p +   \ell_x^2  \,  \frac{p}{x}    +  \ell_p^2  \, \frac{ x}{p}, \qquad   D =  (0, \infty) 
\label{15}
\eeq
which, as the model ${\rm II}$,  is symmetric under the interchange $x-p$, differing from
it in the term $\ell_x^2 \ell_p^2/xp$ in the Hamiltonian. The map (\ref{13}) becomes
in this case

\beq
x' - \ell'_x  = \int_0^x dy \, \frac{ \ell_p   y }{\sqrt{  y^2  + \ell_x^2}}  \longrightarrow x' = \ell_p  \sqrt{ x^2 + \ell_x^2}, \quad \ell'_x = \ell_x \ell_p,
\label{16}
\eeq
which plugged into (\ref{14}) yields 

\beq
w'_{III}(x') =  \ell_p   \sqrt{  x^2 +  \ell_x^2}    =  x'.
\label{13d}
\eeq
so  that this model actually coincides with 
 the model ${\rm I}$ defined in (\ref{10}). 

In the definition of the family of Hamiltonians (\ref{1}) we have imposed the positivity condition
on  $U(x)$ and $V(x)$. Let us suppose for a while that $V(x) =0$. One can see that
a reparametrization $x \rightarrow x'$ can  bring $U(x)$ to  $x'$ and so,  all the Hamiltonians of the form
$U(x) p$, are equivalent to $xp$.

\section{Lagrangian formulation: relativistic spacetime picture}
\label{sec:relativity}

An essential  feature of General Relativity is that the fundamental equations of the theory take the same form
in all coordinate systems. As shown in the previous section, this is also a feature
of the models defined by the Hamiltonians (\ref{1}),  with respect to the coordinate $x$. 
Henceforth, one may suspect the existence of a general relativistic theory lying  behind the models  (\ref{1}),
which would provide them  with a  spacetime interpretation. In this section we shall show  that this is indeed
the case via  the Lagrangian formulation.   

The Lagrangian  associated to  the Hamiltonian (\ref{1}) is given by

\beq
L = p \, \dot{x}  -  H  = p \, \dot{x} -  U(x) p - \frac{ V(x)}{p}. 
\label{g1}
\eeq
In standard classical mechanics,  the Lagrangian can be  expressed solely   in terms of  $x$ and $\dot{x}= dx/dt$,
as $L = m \dot{x}^2/2 - V(x)$,  where  $m$ is  the mass of the particle and $V(x)$ is the potential.  
To find $L(x, \dot{x})$ in our case,   we  use the Hamilton equation of motion 

\beq
\dot{x} = \frac{ \partial H}{ \partial p} = U(x) - \frac{ V(x)}{p^2}, 
\label{g1b}
\eeq
to  eliminate  $p$ in terms of $x$ and $\dot{x}$. This gives two solutions  

\beq
p = \eta \sqrt{ \frac{V(x)}{ U(x) - \dot{x}}}  \, , \qquad \eta = \pm 1,  
\label{g2}
\eeq
that depend on the sign of the momenta, $\eta = {\rm sign} \, p$, which is a conserved quantity. 
The positivity of $U(x)$ and $V(x)$, imply 
that the velocity $\dot{x}$ must never exceed  the value of $U(x)$, for the momentum not to become an imaginary number. 
Substituting (\ref{g2}) into (\ref{g1}),  yields a  Lagrangian, 

\beq
L_\eta(x, \dot{x})  = - 2 \eta \sqrt{ V(x) ( U(x) - \dot{x})  }.
\label{g3}
\eeq
for each value of $\eta$. 
Notice that eq.(\ref{g2}) is singular if $V(x)=0$, so  that 
the  Lagrangian cannot be expressed in terms of $x$ and $\dot{x}$. This is precisely the situation of the
usual  $xp$ Hamiltonian, whose Lagrangian, $L = p (\dot{x} - x)$,  has to be considered as a function
of the three variables $x, \dot{x}$ and $p$.  Later on we shall give an interpretation of this  peculiar fact.

At the classical level we can restrict the motion of the particle to a definite value of $\eta$,
but not  at the quantum level, where both signs would be required. 
The action $S$ corresponding to (\ref{g3}) is ( we choose $\eta = 1$ )

\beq
S   =  \int \, dt  \, L_1 = - 2   \int  \sqrt{ U(x) V(x) \,  (dt)^2 - V (x)  dt \, dx}, 
\label{g5}
\eeq
and it coincides with  the action of a particle moving in 1+1 dimensional
spacetime with metric $g_{ \mu \nu}$, i.e

\beq
S =  -  \int d \sigma \sqrt{ - g_{\mu \nu} \frac{ d x^\mu}{ d \sigma}  \frac{ d x^\nu}{ d \sigma} } = - 
\int \sqrt{ - g_{\mu \nu} d x^\mu d x^\nu }, 
\label{g6}
\eeq
where $\sigma$ parametrizes the worldline( we have set the mass of the particle to 1).  
Making the identifications

\beq
x^0 = t, \qquad x^1 = x, 
\label{g7}
\eeq
the metric tensor becomes

\beq
g_{00} =  - 4 U(x) V(x) \equiv -  4 W(x), \qquad g_{01} = g_{10} = 2  V(x), \qquad g_{11} = 0. 
\label{g8}
\eeq 
In our conventions,   the square of a  line element $dx^\mu$ will be  defined as 

\beq
(ds)^2 = g_{\mu \nu} d x^\mu d x ^\nu, 
\label{g9}
\eeq
so that a time-like distance corresponds to $(ds)^2 < 0$, and a space-like distance to $(d s)^2 > 0$. 
Eq.(\ref{g8}) imply  that,  under general transformations of the coordinate $x$, the function $W(x)$ is
a scalar,  while the $V(x)$ is a cotangent vector, 
in agreement with the results of section \ref{sec:classical}.  The determinant
of the metric (\ref{g8}), given by

\beq
g  \equiv \det \, g_{ \mu \nu} =  - 4 V^2(x),  
\label{g10}
\eeq 
implies  that $g_{\mu \nu}$ is a  non degenerate  Minkowski metric since  $V(x)>0$. 

To gain further insight into  the spacetime structure underlying the $xp$ models,  we shall
employ  the light-cone formalism, which we pass now to describe. 
Any two dimensional  metric  is conformally equivalent to a flat metric. This means that it can be written as

\beq
(ds)^2 = e^{ 2 \chi} \, dx^+   dx^-, \qquad  \chi = \chi(x^+, x^-),  
\label{g12}
\eeq
where $x^\pm$ are the  light-cone variables and $e^\chi$ is  the conformal factor. To find the transformation
from the variables $x^0, x^1$ to the light-cone variables $x^+, x^-$, we use the 
transformation law of the metric tensor

\beq
g'_{\alpha, \beta}(x') = \frac{ \partial x^\mu}{ \partial x'^\alpha}  \frac{ \partial x^\nu}{ \partial x'^\beta}  \;  g_{\mu \nu}(x),   
\label{g13}
\eeq
where the  ligh-cone  metric corresponds to 

\beq
g_{++} = g_{--} =0, \qquad g_{+-} = g_{-+} = \frac{1}{2} e^{2 \chi}. 
\label{g14}
\eeq
Eqs. (\ref{g8}) and (\ref{g14}), allow us to write (\ref{g13})  as

\barray
0 & = & \partial_+ x^0 \left[  W(x^1) \partial_+ x^0 - V(x^1) \partial_+ x^1 \right] 
\label{g15a} \\
0 & = & \partial_- x^0 \left[  W(x^1) \partial_- x^0 - V(x^1) \partial_-  x^1 \right] 
\label{g15b}  \\
e^\chi & = &  - 8 W(x^1) \partial_+ x^0 \partial_- x^0 + 4 V(x^1) \left[ \partial_+ x^0 \partial_- x^1 + \partial_+ x^1 \partial_- x^0  \right].  
\label{g15c}
\earray 
Let us suppose, for a while,  that $x^0$ depends non trivially on $x^+$ and $x^-$, i.e.
$ \partial_\pm  x^0 \neq 0$. Hence eqs. (\ref{g15a}) and (\ref{g15b}),  would imply

\beq
\partial_\pm  x^0 = \frac{ \partial_\pm x^1}{U(x^1)} = \partial_\pm \int^{x^1}_{\ell_x}  \frac{dy}{U(y)} \Longrightarrow
x^0 =  \int^{x^1}_{\ell_x}  \frac{dy}{U(y)}  + {\rm cte}, 
\label{g15d}
\eeq
so that $x^0$ would be a function of $x^1$, which is  a contradiction since they are independent
variables.  We shall make the choice that $x^0$  only depends  on $x^+$

\beq
x^0 = f(x^+). 
\label{g16}
\eeq
Eq.(\ref{g15b}) is fulfilled automatically,  and eq.(\ref{g15a}) becomes

\beq
\partial_+  x^0 = \frac{ \partial_+ x^1}{U(x^1)} = \partial_+ \int^{x^1}_{\ell_x} \frac{dy}{U(y)} \Longrightarrow
x^0 =  \int^{x^1}_{\ell_x}  \frac{dy}{U(y)}  - g(x^-), 
\label{g17}
\eeq
and so 

\beq
\int^{x^1}_{\ell_x}  \frac{dy}{ U(y)} = f(x^+) + g(x^-), 
\label{g18}
\eeq
where $f(x^+)$ and $g(x^-)$ are generic functions. $x^1(x^+, x^-)$ is given,  in an implicit
way,  by eq.(\ref{g18}).  Finally, eqs.(\ref{g15c}) and (\ref{g18})  provides  the conformal
factor,

\beq
e^{2 \chi}   = 4 W(x^1) \partial_+ f(x^+) \partial_- g(x^-). 
\label{g19}
\eeq
Equations (\ref{g16}) and (\ref{g18}) give the  map 
 from $x^{0,1}$ to $x^{\pm}$. However the map is not unique due to the freedom 
 in choosing  $f(x^+)$ and $g(x^-)$. This simply reflects the invariance of the metric
 (\ref{g12}) under general conformal transformations, 
  $x^+ \rightarrow \tilde{f}(x^+)$ and $x^- \rightarrow \tilde{g}(x^-)$. 
  
In the conformal gauge  (\ref{g14}),  the   tensors and connections  simplify considerably. 
The Christoffel 
symbols, $\Gamma_{\mu \nu}^\lambda$,  have only $\pm$ non vanishing components 

\beq
\Gamma^+_{ ++} = g^{+-} \partial_+ g_{+ -} = 2 \partial_+ \chi, \qquad  
\Gamma^-_{ --} = g^{+-} \partial_- g_{+ -} = 2 \partial_- \chi, 
\label{g20}
\eeq
so  that the equations  of the  geodesics $x^\pm(s)$ read  

\beq
\frac{ d^2 x^\pm }{ d s^2} + 2  \left( \frac{ d  x^\pm }{ d s} \right)^2  \;  \partial_\pm   \chi =0, 
\label{g20b}
\eeq
where $s$ is the propertime, i.e. $(ds)^2 = - e^{ 2 \chi}  dx^+ dx^-$. 
The Ricci tensor, $R_{\mu \nu}$,  becomes

\beq
R_{+ - } =  R_{- +} = - 2 \partial_+ \partial_- \chi, \qquad R_{++} = R_{--} = 0, 
\label{g21}
\eeq
and the Ricci scalar,  $R= g^{\mu \nu} R_{\mu \nu}$, 

\beq
R = - 8 e^{ - 2 \chi} \partial_+ \partial_- \chi.
\label{g22}
\eeq
It is not difficult to show that  

\beq
R(x)  = - \frac{ 1}{ V(x)} \partial_x  \left[ \frac{\partial_x W(x)}{ V(x)} \right]. 
\label{g24}
\eeq
In the symmetric gauge (\ref{9c}),  one has $V(x) = w(x)$ and $W(x)= w^2(x)$, so (\ref{g24})
becomes

\beq
 R(x)  =   - 2  \frac{ \partial_x^2 w(x) }{ w(x)}  \label{g25a}
 \label{g25} 
\eeq
We shall use this formula  to relate different $xp$ models to the underlying 
 spacetime geometries. 

\vspace{0.4 cm}

{\bf Flat spacetimes}

\vspace{0.4 cm}


In flat spacetimes  the scalar curvature vanishes. Equation (\ref{g25}) provides the function
$w(x)$ corresponding to  these cases

\beq
R(x) = 0, \quad \forall x \in D \Longleftrightarrow w(x) = \alpha  x  + c,    \qquad \alpha  \geq   0.
\label{g26}
\eeq
The condition $\alpha \geq 0$ comes from the positivity of $w(x)$.  If $\alpha  >0$,  the shift 
$x  \rightarrow x - c/\alpha$,  brings $w(x)$ to the form

\beq
 w(x) = \alpha   x,    \qquad \alpha  >    0, \qquad D = (h, \infty). 
\label{g261}
\eeq
For $\alpha = 1$, this model coincides with  (\ref{10}). The 
value of $w_0 \equiv  w(h)= \alpha h$,
is independent of reparametrizations. 
If $\alpha = 0$,  $w(x)$  is constant and the Hamiltonian  is simply  

\beq
w(x) = c  > 0,  \qquad H = c  \left( p + \frac{1}{p} \right),  \qquad D = (0, \infty ),
\label{g264}
\eeq
where we have choosen the origin as the boundary of the interval $D$
(see \ref{sec:kondo}).

\vspace{0.4 cm}

{\bf Berry-Keating model}

\vspace{0.4 cm}

This model was defined in  eq.(\ref{10b}). The spacetime has a scalar curvature 

\beq
w(x)  = x + \frac{h^2}{x} \Longrightarrow R(x) = - \frac{4 h^2}{ x^2 (  x^2 + h^2)}
\label{g28}
\eeq
which is always negative,  vanishes asymptotically as $x^{-4}$, and  diverges  at the origin as $x^{-2}$.

\vspace{0.4 cm}

{\bf Spacetimes with constant negative curvature}

\vspace{0.4 cm}

Eq.(\ref{g25a}) admits  a solution  with constant negative curvature

\beq
R(x)  = - |R|,  \quad    \forall x \in D    \Longleftrightarrow w(x) = w_0     
 \cosh( x  \sqrt{ |R|/2}    ),  \quad   w_0 >0
\label{g27}
 \eeq
where $w_0 >0$ to guarantee  the positivity of $w(x)$. We shall take $D=(0, \infty)$. 
There  is 
also a solution of (\ref{g25a})  with positive curvature  involving  
the cosine function,  but it requires finite 
 $D$  domains  in order to maintain the positivity of $w(x)$. 
 We shall not consider this case  below. 
The interest of solution (\ref{g27}) is that the semiclassical spectrum 
coincides with that of the harmonic oscillator (see section \ref{sec:semiclassical}). 

\vspace{0.4 cm}

{\bf Linear-log model}

\vspace{0.4 cm}

An interesting  variation of the linear potential (\ref{g26}) is to add a subleading logarithmic term, i.e.

\beq
w(x) = \alpha  x + \beta  \log x \Longrightarrow R(x) = \frac{ 2 \beta }{ x^2 ( \alpha  x  + \beta  \log x)} 
\rightarrow  \frac{ 2 \beta}{ \alpha  x^3}, 
\label{g29}
\eeq
where  the curvature  decays asymptotically as $x^{-3}$, with a sign determined by that of $\beta$ ($\alpha >0$).

\vspace{0.4 cm}

{\bf Power like models}

\vspace{0.4 cm}

These models are defined by

\beq
w(x)  = A  \,   x^\alpha  \quad (A, \alpha > 0)   \Longleftrightarrow  R(x) = - \frac{ 2  \alpha (\alpha -1)}{ x^2}. 
\label{g29b}
\eeq
We impose the condition $\alpha>0$ to have a monotonic  increasing function $w(x)$.  If $\alpha \neq 1$, 
the curvature vanishes asymptotically as $x^{-2}$, and its sign is negative for $\alpha > 1$ and positive
for $0 < \alpha < 1$.  In section \ref{sec:semiclassical} we shall show that the asymptotic behavior of the curvature 
 is intimately related to the semiclassical spectrum of the model.

\subsection{Flat spacetimes} 
\label{subsec:flat}

Let us  study in more  detail the model (\ref{g261}).  Since the curvature vanishes, 
 there is   a choice of $f(x^+)$ and $g(x^-)$ 
 for which  the conformal factor $e^{ \chi}$  is  constant,  and therefore  the geodesics
 are  straight lines, it is given by 

\beq
f(z) = g(z) = \frac{1}{2 \alpha} \log z,
\label{g32}
\eeq
which plugged into eqs.(\ref{g16}) and (\ref{g18}) yields  ($h \equiv \ell_x$)

\beq
x^0 = \frac{1}{2 \alpha} \log x^+, \qquad x^1 = h (x^+ x^-)^{1/ 2 }, 
\label{g33}
\eeq
and conformal factor  (recall (\ref{g19}))

\beq
e^\chi  = h.  
\label{g35}
\eeq
The line element  

\beq
(ds)^2 = h^2dx^+ dx^-, 
\label{g36}
\eeq
implies that   the geodesics are straight lines in the  $x^\pm$ plane. Not the whole $x^\pm$   plane
is available for the motion of the particle 
because it  is constrainted to the interval $D= (h,\infty)$. In light-cone coordinates 
the spacetime domain,  ${\cal U}$, can be  obtained from eq.(\ref{g33}) 

\beq
{\cal U} = \{ (x^+, x^-) \, | \,    x^\pm  \in (0, \infty), \quad x^+ x^- \geq 1 \}. 
\label{g37}
\eeq
If    $x^+$ and $x^-$ denote   the vertical and horizontal axes of the plane, then 
${\cal U}$  is the region in the first quadrant  that is 
above the hyperbola $x^+ x^- = 1$.
This hyperbola is  the worldline of the point $x^1=h$. 
More generally, the worldlines of any point $x^1 \geq h$, are given by 
the hyperbolas  $x^+ x^- = (x^1/h)^{2}$. $x^+$
is  the light-cone time coordinate  and it  flows upwards. Eliminating $x^+$ in   eqs.(\ref{g33}) one finds

\begin{figure}
\begin{indented}
\item[]\includegraphics[width=.45\linewidth]{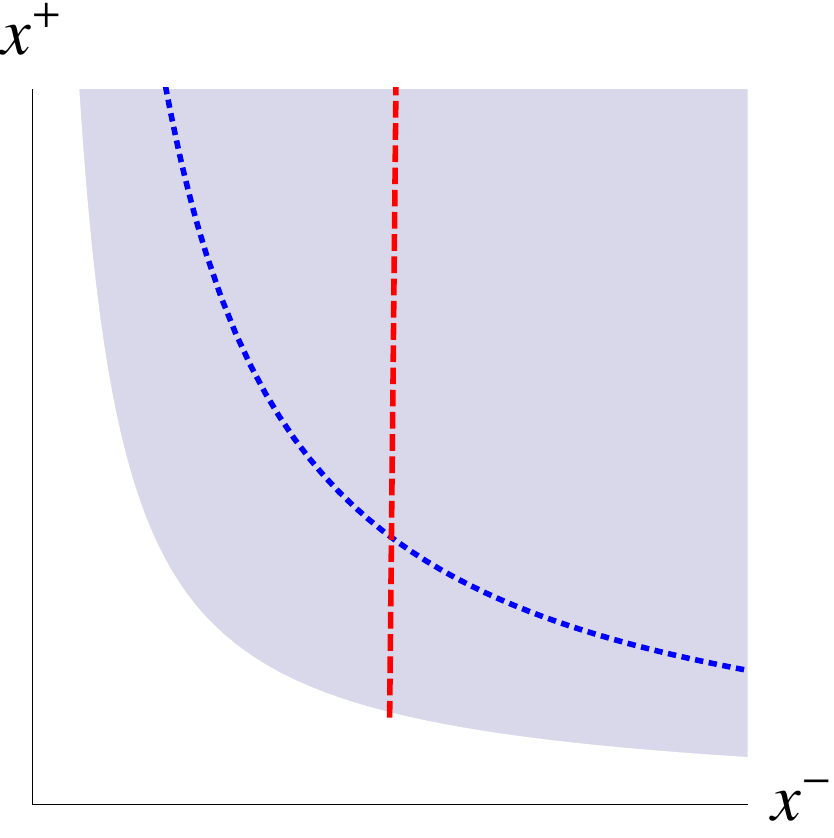}
\hspace{1cm}
\includegraphics[width=.45\linewidth]{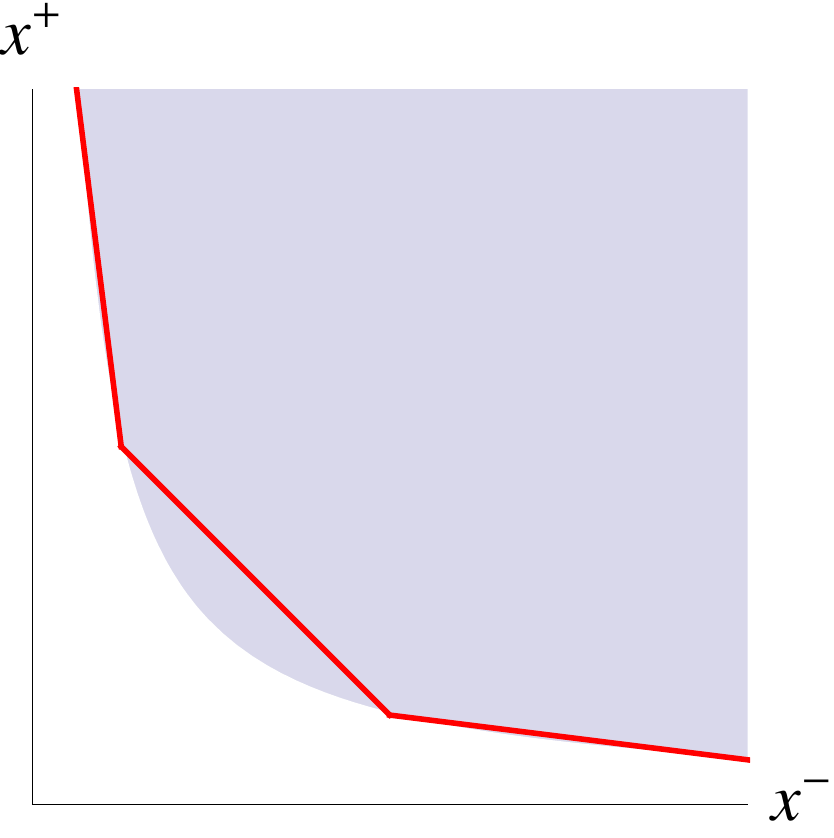}
\end{indented}
\caption{The region in shadow  represents the universe ${\cal U}$ in light-cone variables defined in  eq.(\ref{g37}). Left:
the hyperbola (dotted line)  corresponds to the worldline of a given position $x^1 = \sqrt{ x^+ x^-}= {\rm cte}$, 
and the vertical (dashed)  line  corresponds to a light ray emanating at the boundary  (eqs. (\ref{g38})). Right:
the worldline of a particle with constants energy $E$, which bounces off at the boundary. 
}
\label{light-cone}
\end{figure}

\beq
x^1 = h  e^{\alpha x^0} \, (x^-)^{1/2}. 
\label{g38}
\eeq
Hence,  the vertical lines, i.e. constant values of  $x^-$, coincide with the classical solutions of the Hamiltonian, 
$\alpha x p$, namely 
$x \propto e^{ \alpha t}$. The line element (\ref{g36}) vanishes
along these trajectories, which therefore
represent light rays  that start at a 
 point on the boundary $x^1 =h$ and scape to infinity as  $x^0 \rightarrow \infty$ (see fig. \ref{light-cone}):

\beq
x^- = {\rm cte} \leftrightarrow {\rm light} \; {\rm ray}. 
\label{g39}
\eeq
The line element (\ref{g36}) also vanishes along the horizontal lines, $x^+ = {\rm cte}$, but they do not correspond 
to light rays since the time coordinate $x^0$ is frozen.  In this theory, the light rays are right movers.  The left moving light rays are absent. 
This chirality is a reflection of  the time reversal symmetry  breaking of the Hamiltonian (\ref{1}).

The causal cone,  i.e. $( d s)^2 < 0$, at each point of ${\cal U}$, is given by the 
second and fourth quadrants, which correspond respectively to the future and past events relative
to that point.  A particle follows straight lines, with negative slope that start and end
at the boundary $x^1 = h$. To show this fact explicitely,  we  solve   the classical  equations of motion 
of the Hamiltonian with $w(x) = \alpha x$,   for   positive energy $E$  (see section \ref{sec:trajectory})

\beq
x^{ 2 }  =  \frac{E}{\alpha} e^{ 2 \alpha(t - t_0)}  - e^{ 4 \alpha ( t - t_0)}, 
\qquad 
p^{ 2}  = \frac{E}{\alpha}   e^{ - 2  \alpha( t - t_0)} -1 , \quad E> 0. 
\label{g40}
\eeq
In light-cone variables (\ref{g33}) this equation becomes  a  straight line 

\beq
q^{-\alpha} \, x^+ + q^\alpha  \, x^- = \frac{E}{w_0}, \qquad E> 0, \quad q= e^{ 2 t_0}  h^{1/\alpha}> 0, \quad
w_0 = h \alpha, 
\label{g41}
\eeq
where $q$ parametrizes the slope that depends on the time $t_0$ where $x(t_0) = |p(t_0)|$. This line 
ends and starts at the points $A$ and $B$ of the boundary $x^+ x^- =1$, with  coordinates

\barray
x^+_{A, B}  = \frac{1}{ x^-_{A,B} }  &  = & \frac{  q^\alpha}{ 2 w_0}    \left( E  \pm  \sqrt{ E^2 - 4 w^2_0} \right)  = q^\alpha \,  
e^{\pm \alpha  \varepsilon},  
\label{g42} 
\earray
where

\beq
 \cosh( \alpha  \varepsilon) \equiv \frac{ E}{ 2 w_0}   \geq 1. 
\label{g43}
\eeq
The energy  $E$ of the classical orbits are bounded by $2 w_0$. 
The entire  worldline of a particle with energy $E$,  is given by a polygonal line made of  linear segments (\ref{g41}), that  
come  from the horizontal axis,  $x^0 \rightarrow - \infty$,   and approaches  the vertical axis, 
 $x^0 \rightarrow + \infty$ (see fig. \ref{light-cone}).
The value of  $q$, that parametrizes each segment,  can be found  matching the initial and final positions of
consecutive segments, i.e.  

\beq
x^+_{A_{n-1}}  = x^+_{B_ n}  \rightarrow q_n = e^{ 2 \varepsilon} q_{n-1}, \qquad n = - \infty, \infty
\label{g44}
\eeq
which  means that the particle in the $(n-1)^{\rm th}$ segment bounces off at $x^1=h$ 
and starts a new orbit corresponding to the $n^{\rm th}$ segment. This polygonal worldline
represents a periodic motion, since after a shift $x^0 \rightarrow x^0 + T_E$, i.e. $x^+ \rightarrow e^{ 2 \alpha  T_E}  x^+$,
the equation (\ref{g41}) remains invariant if $q \rightarrow e^{ 2 T_E} \, q$ which, according to (\ref{g44}),
gives the period $T_E$ as a function of the energy, i.e

\beq
T_E = \varepsilon  = \frac{1}{\alpha}  \cosh^{-1} \frac{ E}{2 w_0}. 
\label{g45}
\eeq

\begin{figure}
\begin{indented}
\item[]\includegraphics[width=.45\linewidth]{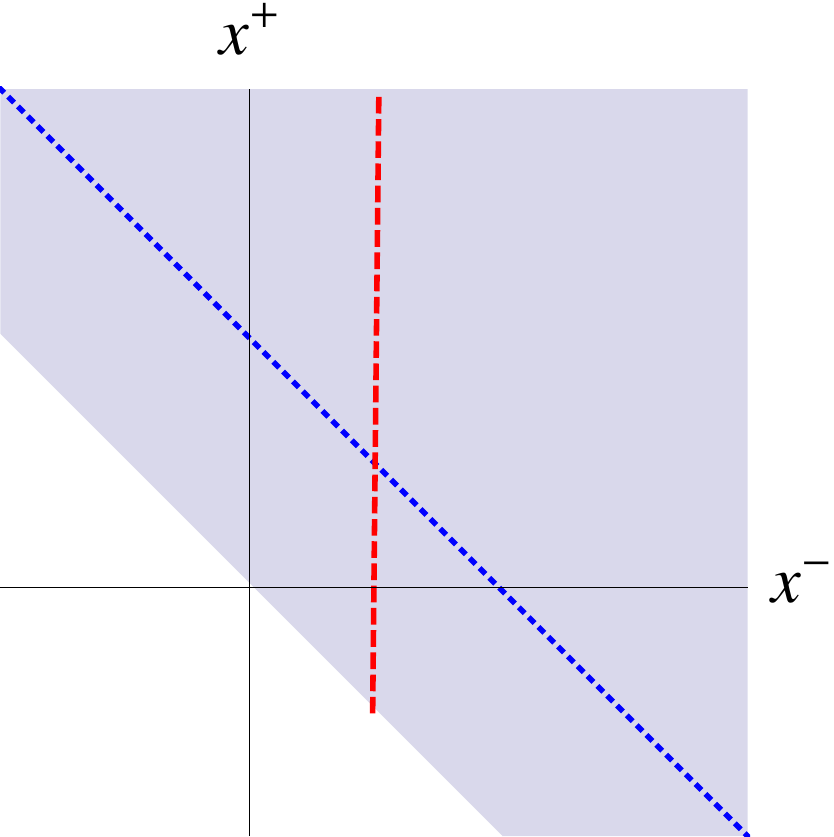}
\hspace{1 cm} 
\includegraphics[width=.45\linewidth]{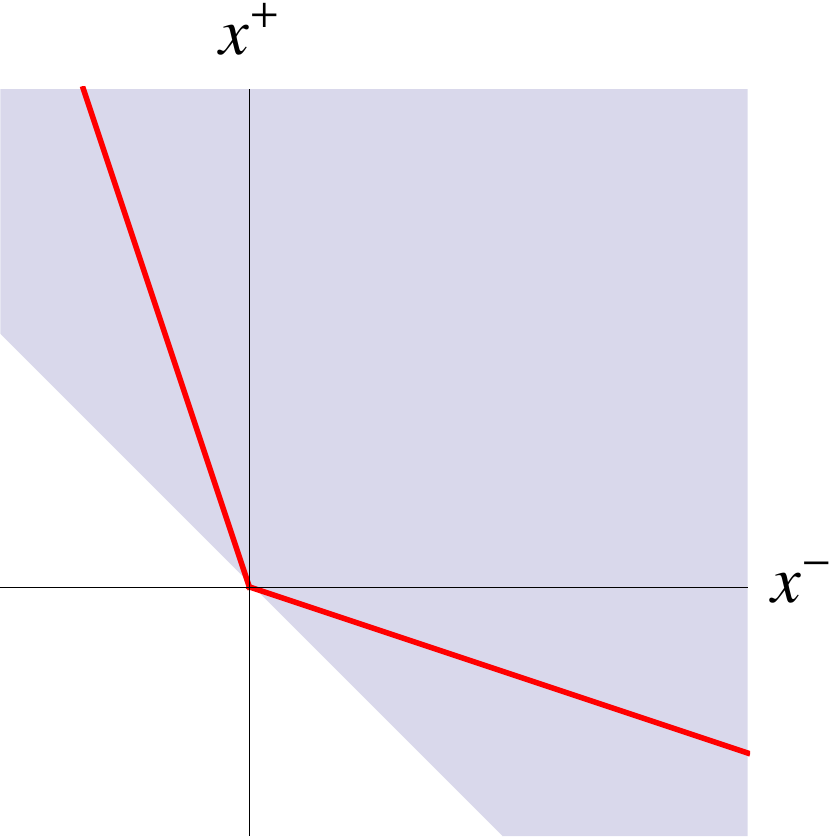}
\end{indented} 
\caption{The region in shadow  represents the universe ${\cal U}$ in lightcone variables defined in  eq.(\ref{g49}). Left:
the hyperbola (dotted  line)  corresponds to the world line of a given position $x^1 = c(  x^+ +  x^-)= {\rm cte}$, 
and the vertical (dashed)  line  corresponds to a light ray emanating at the boundary $x^+ + x^-=0$. Right:
 worldline of a particle with constants energy $E$, describe by eqs.(\ref{g52a}) and (\ref{g52b}) with $t_0=0$. 
}
\label{light-cone2}
\end{figure}

Let us next study the model defined in equation (\ref{g264}), which also has 
a vanishing curvature. A choice that leads to a constant conformal factor is 

\beq
f(z) = g(z) = z, 
\label{g46}
\eeq
which using  eqs.(\ref{g16}) and (\ref{g18}) yields 

\beq
x^0 =  x^+, \qquad x^1 = c (x^+ + x^-), 
\label{g47}
\eeq
and 

\beq
e^\chi = 2  c.  
\label{g48}
\eeq
The constraint $x^1 \geq 0$ provides  the domain of spacetime  
\beq
{\cal U} = \{ (x^+, x^-) \, |  \,   x^+ +  x^- \geq 0 \}. 
\label{g49}
\eeq
which is depicted in fig.\ref{light-cone2},  which also shows the light-rays and the worldline of the points. 
The classical equations of motion have  the solutions (we choose $E>0$)

\barray
x=  c (t-t_0) ( 1 - e^\varepsilon),  & & \qquad  p = e^\varepsilon,  \label{g50a} \\
x=  c (t-t_0) ( 1 - e^{-\varepsilon}),  & & \qquad  p = e^{-\varepsilon},  \label{g50b}
\earray
where $\varepsilon$ is defined as

\beq
\cosh \varepsilon = \frac{ E}{2 c}, \qquad \varepsilon \geq 0. 
\label{g51}
\eeq
The solution (\ref{g50a}) describes  the particle moving towards the origin, i.e. $\dot{x} <0$, which
it is reached at $t= t_0$. At that moment it  bounces off and  starts to move to the right, i.e. $\dot{x} >0$,
as described by eq.(\ref{g50b}).  In the lightcone coordinates (\ref{g47}), eqs.(\ref{g50a}, \ref{g50b})
becomes

\barray 
e^{- \varepsilon/2} \, x^+ + e^{\varepsilon/2} \, x^- & = &  - 2 t_0 \sinh( \varepsilon/2),  \label{g52a} \\
e^{ \varepsilon/2} \, x^+ + e^{\varepsilon/2} \, x^- & = &  - 2 t_0 \sinh( \varepsilon/2).  \label{g52b}
\earray
Fig. \ref{light-cone2} illustrates the form of these  trajectories.

A conclusion of the results we obtained in this subsection  is that the scalar curvature by itself
does not fully characterize a model and that the boundary of spacetime may play an essential role.

Before we leave this section we want to make some   remarks: 

\begin{itemize}

\item  The reformulation of the Hamiltonians (\ref{1}) as  relativistic models described by the Lagrangians
(\ref{g3}) holds strictely speaking at the classical level. At the quantum level, this relation is more
involved. Indeed, one should start from the path integral in phase space and perform the  integration
over the variable $p$. For Hamiltonians of the form $H = p^2/2m + V(x)$, this integral is gaussian and 
one gets the familiar Feymann path integral with Lagrangian $L = m \dot{x}^2/2- V(x)$. However for the Hamiltonians (\ref{1}), 
 the integration over $p$ is not gaussian and in the saddle approximation, 
 one gets extra terms in addition to the Lagrangian (\ref{g3}).
 In any case, the quantization of these models will be done in section \ref{sec:quantization} 
 using  the Hamiltonian formulation, where  this issue  does not arise.

\item Given a model in the symmetric gauge (\ref{9c}), one can find a  new coordinate $x'$ such that  $U(x') =1$, that we 
shall call the $p-$gauge.  The transformation $x' = f(x)$,  and the new function $V'(x') \equiv V_p(x')$,  can be derived from eq.(\ref{9}) and the scalar nature 
 of  $W(x)$

\beq
x' - \ell'_x = f(x) = \int_{\ell_x}^x \frac{ dy}{w(y)}, \qquad V_p(x') = w^2(f^{-1}(x')). 
\label{pg}
\eeq
In the $p-$gauge the Hamiltonian takes the canonical form

\beq
H = p + \frac{ V_p(x)}{p}, 
\label{pgb}
\eeq
whose square 

\beq
H^2 = p^2 + 2 V_p^2(x) + \frac{ V_p^2(x)}{p^2}, 
\label{pgc}
\eeq
coincides with the standard Hamiltonian with  a positive potential $2 V_p^2(x)$,  except for the $V_p^2(x)/p^2$ term. 
One finds for example that the functions $V_p(x)$ associated to the models (\ref{10}) and (\ref{10b}) are given by

\barray
H_{\rm I} &  \rightarrow & V_{p, \rm I}(x) = h^2 \, e^{ 2 x} , \hspace{1.6cm}  D = (0, \infty), \label{pg1} \\
H_{\rm II} &  \rightarrow & V_{p, \rm II}(x) = \frac{ h^2 e^{2 x}}{ 1 - e^{- 2 x}} ,  \qquad D = (0, \infty).   \label{pg2}
\earray 
which seem to have some relation with the Liouville model and  the Morse potentials studied in  \cite{L09} (see references therein),
whose spectrum is  related  to   the Riemann zeros in average. 

\end{itemize}

\section{Classical trajectories and equations of motion}
\label{sec:trajectory} 

\begin{figure}
\begin{indented}
\item[]\includegraphics[width=.45\linewidth]{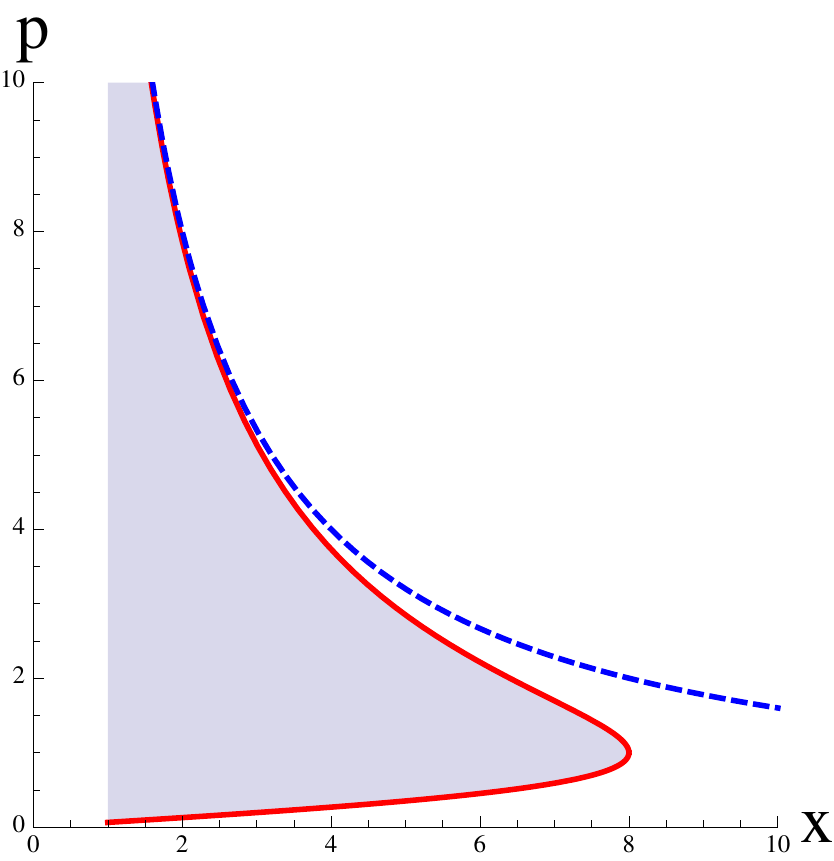}
\hspace{0.4cm}
\includegraphics[width=.45\linewidth]{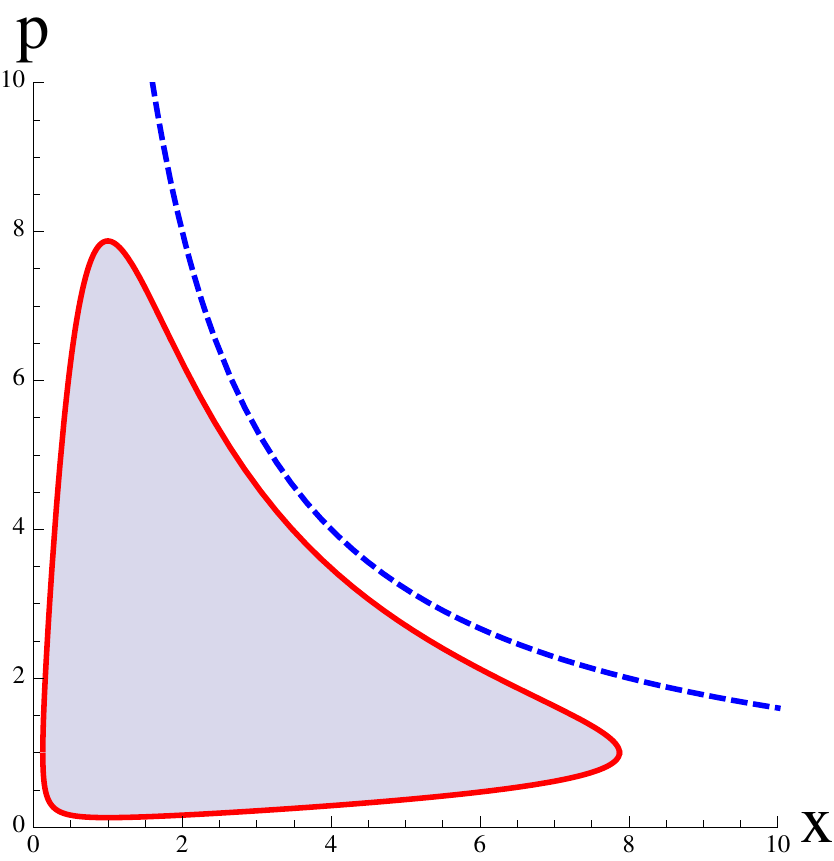}
\end{indented}
\caption{
Classical trajectories   of the Hamiltonians (\ref{1a}) (left) and (\ref{1b}) (right)  in phase space with $E>0$. 
We include for comparison 
the classical trajectory of the $xp$ model,  given by the hyperbola $E = xp$ (dashed lines).. The parameters of the latter
Hamiltonians are choosen as $\ell_x = \ell_p =1$. The classical trajectories with negative energy  can be obtained from the
trajectories with positive energy replacing  $p \rightarrow -p$. 
  }
\label{tray}
\end{figure}

In the symmetric gauge (\ref{9c}),  the classical trajectories in phase space are curves with  constant energy $E$, 

\beq
E = w(x) \left(p + \frac{1}{p} \right).
\label{c1}
\eeq
For each position $x$,  there are in general two different  values of the momentum $p$ that we denote as

\beq
p_\pm(x, E) = \frac{1}{ 2 w(x)} \left( E \pm \sqrt{ E^2 - 4 w^2(x)} \right), 
\label{c2}
\eeq
and   satisfy the relation

\beq
 p_+(x,E) = \frac{1}{p_-(x,E)}.
\label{c2b}
\eeq
In fig. \ref{tray},  we plot the   classical trajectories  
 corresponding to the models  (\ref{10}) and (\ref{10b}).  
In all these models,  the trajectories are clockwise for positive energy and
anticlockwise for negative energy.  This is a consequence of the time reversal breaking of the Hamiltonian (\ref{1}). 
In the model (\ref{10}), the particle starts at 
$x = h$, with a high momentum, say  $p_{\rm high}$.  Then it  moves to the right, while 
its  momentum decreases until  $|p| = 1$, where $x$  reaches a maximum  $x_{\rm M}(E)$. Afterwards,  the particle moves
back  towards  lower values of $x$ and $|p|$. When the particle  reaches the boundary of the interval, i.e. 
$x= h$, with a momenta $|p_{\rm low}| <  1$,  it bounces off acquiring the original  momenta
 $p_{\rm high} =1/p_{\rm low}$, that satisfy eq.(\ref{c2b}). The latter relation can be written as
$\log  |p_{\rm high}|   = - \log  |p_{\rm low}|$ 
which is the analogue  of  the reflection of a particle in a wall. 
After the reflection, the particle restarts the motion from its original position
and momentum, so that the classical trajectory is closed and periodic \cite{SL11}. 

In the model (\ref{10b}), there is no a wall for the motion of the particle which  follows
a continuous and differential orbit  around the point $(h, 1)$ in phase space \cite{BK11}. 
In both   models,
for large values of $x$ and $p$, the classical trajectories approach the hyperbola $E= xp$, but they departure
from it   whenever $x$ or $|p|$ are of the order of the parameters $h$ or $1$.

The Hamilton equations of motion for the general Hamiltonian (\ref{1}) are given by

\beq 
\dot{x} = w(x) \left(1-   \frac{1}{p^2} \right), \qquad \dot{p} = - w'(x) \, \left( p + \frac{1}{p}  \right),  
\label{c5}
\eeq
where $\dot{x} = dx/dt, \;  w'(x) = dw(x)/dx$. Computing $\ddot{x}$,  and expressing  $p$ as a function of $\dot{x}$
and $w(x)$, one finds

\beq
\ddot{x} = w(x) w'(x) \left( - 4 + \frac{6 \dot{x}}{w(x)} -  \frac{ \dot{x}^2}{ w^2(x) }  \right). 
\label{c515}
\eeq
In section \ref{sec:relativity}, we formulated our model as a general  relativistic theory.
Equations  (\ref{g20b}) for the geodesics are therefore expected to
follow  from  eq. (\ref{c515}). 
Indeed, let us  choose for simplicity $f(z) = g(z) = z$, so that eqs.(\ref{g16}),
(\ref{g18}) and (\ref{g19}) become

\beq
t = x^+, \qquad \partial_\pm x = w(x), \qquad e^{2 \chi}  = 4 w^2(x), 
\label{c511}
\eeq
where $x^0 =t$ and $x^1 = x$. 
The  proper time $s$ and  the  time  $t$ along  the trajectory are related by 

\beq
(ds)^2 = - e^{2 \chi}  dx^+ dx^- = 4 w^2(x) \left( (dt)^2 - \frac{dt \,  dx}{ w(x)}  \right), 
\label{c512}
\eeq
so 

\beq
\frac{ d x^+}{ ds} = \frac{1}{2 w(x)} \left( 1- \frac{ \dot{x}}{ w(x)}  \right)^{-1/2}. 
\label{c513}
\eeq
 Taking a derivative respect  to $t$ in this equation yields
 
 \barray
 \frac{ d^2  x^+}{ ds^2} \div  \left( \frac{ d x^+}{ ds}  \right)^2 & = & 
\left( 1- \frac{ \dot{x}}{ w(x)} \right)^{-1} \left[ - \frac{ w'(x) \dot{x}}{ w(x)} + 
\frac{ w'(x) (\dot{x})^2}{ 2 w^2(x)} + \frac{ \ddot{x}}{ 2 w(x)} \right], 
 \label{c514}
 \earray  
 which together with  (\ref{c515}),  leads to

 \beq
 \frac{ d^2  x^+}{ ds^2}+  \left( \frac{ d x^+}{ ds}  \right)^2  2 w'(x) = 0.
 \label{c516}
 \eeq
 This equation  coincides with (\ref{g20b}) (for  $x^+$)  as follows from 

\beq
\partial_+ \chi = \frac{  \partial_+ w}{ w} = \frac{\partial_+ x \;   \partial_x w }{w} =  w'(x).
\label{c517}
\eeq
The geodesic equation  for $x^-$ can be derived in a similar way.

\begin{figure}[t!]
\begin{center}
\includegraphics[width=.35\linewidth]{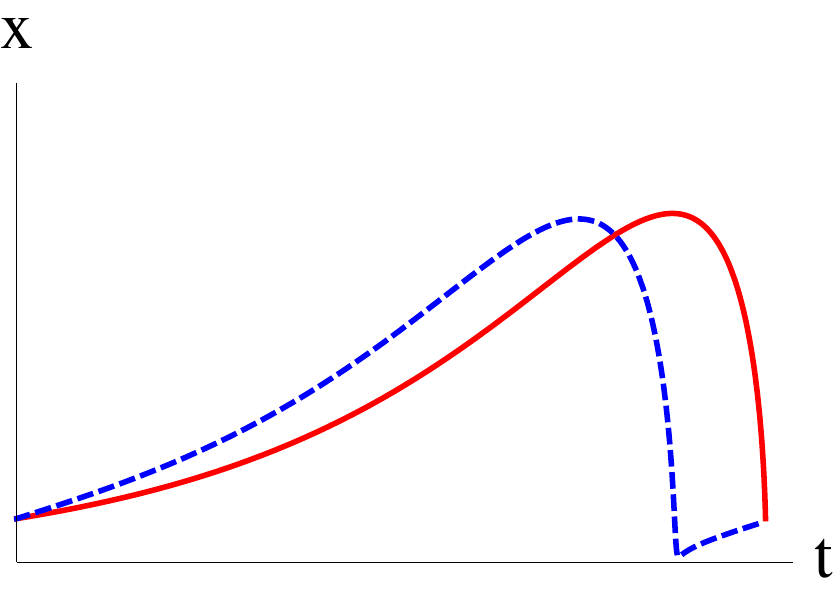}
\hspace{0.6cm}
\includegraphics[width=.35\linewidth]{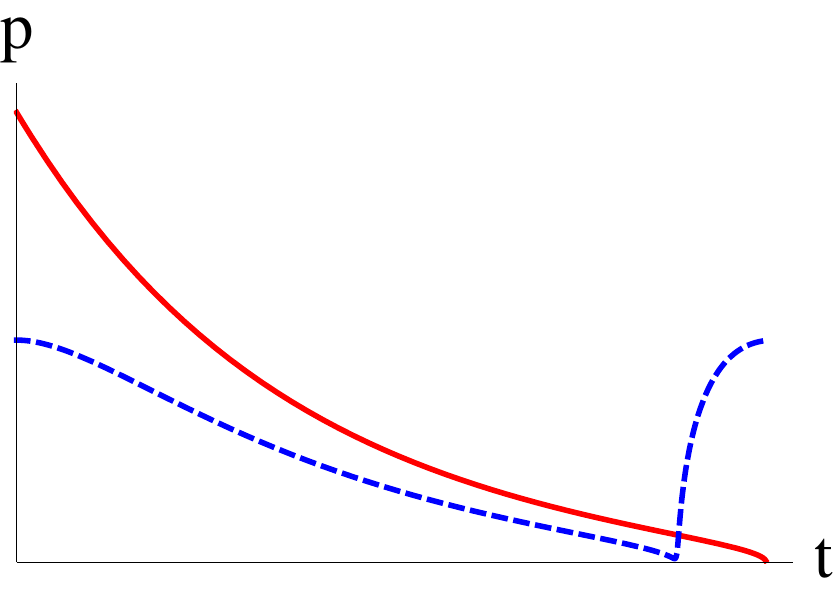}
\end{center}
\caption{
Plot of $x(t)$ and $p(t)$ for the Hamiltonians $H_{\rm I}$ (continuous red curve) and $H_{\rm II}$ (dotted blue curve).
  }
\label{xpt}
\end{figure}

In the model (\ref{g261}) the solution of the eqs. of motion (\ref{c5})
is given by 

\beq
x^{ 2 }  =  \frac{E}{\alpha} e^{ 2 \alpha(t - t_0)}  - e^{ 4 \alpha ( t - t_0)}, 
\qquad 
p^{ 2}  = \frac{E}{\alpha}   e^{ - 2  \alpha( t - t_0)} -1 , \quad E> 0, 
\label{g40b}
\eeq
where $t_0$ is the instant where $x= p$. At the initial and final  times, $t_{i, f}$
the particle is at $x = h$, which  implies 

\beq
e^{ 2 \alpha ( t_{f,i} - t_0)} = \frac{ E}{ 2 \alpha} \pm 
\sqrt{ \left(  \frac{E}{ 2 \alpha} \right)^2 - h^2}, 
\label{g41b}
\eeq
so that the period of the trajectory coincides with (\ref{g45}), 

\beq
T_E = t_f - t_i = \frac{ 1}{ \alpha} \cosh^{-1} \frac{ E}{ 2 w_0} \,  \rightarrow \,  \frac{1}{\alpha} \log \frac{ E}{ w_0}, 
\quad ( E>> w_0). 
\label{g42b}
\eeq
The trajectories of the Berry-Keatig model  with Hamiltonian $H_{\rm II}$,
 (recall eq. (\ref{10b})), are given in terms of elliptic functions \cite{BK11}. 
In fig \ref{xpt}, we plot $x(t)$ and $p(t)$
for the Hamiltonians $H_{\rm I}$ and $H_{\rm II}$ with the same energy. 
The discontinuity of the curve for the $H_{\rm I}$ model is in contrast  with its  continuity for the 
$H_{\rm II}$ model. Despite of this fact,  both curves have almost the same period. 
It is  remarkable how fast the particle retraces its trajectory near the origin, which is the reason  why
the periods $T_E$ in the two models converge for large trajectories.

The general expression  of the period of a trajectory with energy $E$, can be obtained integrating eqs.(\ref{c5}) 


\beq
T_E =  \int_{x_{\rm m}}^{x_{\rm M}}  \frac{ dx}{w(x)} \frac{ E}{ \sqrt{ E^2 - 4 w^2(x)}}, \qquad 
E = 2 w(x_{\rm m}) = 2 w(x_{\rm M}), 
\label{33}
\eeq
where $x_m$ and $x_M$ are the turning points.
The minimal value $x_m$ can actually coincide with the boundary value of the interval $D$,
as for the $H_{\rm I}$ model.

\section{Semiclassical analysis}
\label{sec:semiclassical}

The number of semiclassical states  with energy between 0 and $E>0$, denoted as  $n(E)$, 
is given by the area in phase space swept  by the  closed trajectory, measured in units of the Planck constant
$2 \pi \hbar$. In the symmetric gauge it reads 

\beq
n(E)   = \frac{1}{2 \pi \hbar}   \int_{0 < H < E} dx \, d p = 
\frac{ 1}{ 2 \pi \hbar} 
 \int_{x_{\rm m}}^{x_{\rm M}}  \frac{dx}{ w(x)}  \,   \sqrt{ E^2 - 4 w^2(x)}, 
\label{38}
\eeq
where $x_{m,M}$ are the turning points of the trajectory (recall  (\ref{33})). We have not included
in (\ref{38}) the Maslow phase. The counting formula for the negative energy states 
is also given by  (\ref{38}) with $|E| = 2 w(x_{m,M})$. In the next section we shall see that this symmery is broken in general
by the quantum model.  
In other gauges, the corresponding formula 
for $n(E)$ is obtained from (\ref{38}),  by the replacements $dx /w(x) \rightarrow dx/U(x)$ and $w^2(x) \rightarrow U(x) V(x)$,
which shows  that $n(E)$ is invariant under reparametrizations of $x$ (see section \ref{sec:classical}). 
The  derivative of the area of the trajectory, $ 2 \pi \hbar \,  n(E)$, respect to $E$,  gives   the period  
  (\ref{33}). In the cases where $w(x)$ is an invertible function, and assuming
  that $x_{\rm m} = x_0$, one can invert eq.(\ref{38}) in order to find the function $w(x)$, or rather
  $x(w)$, that  produces a given  $n(E)$. The formula is given
  by  (see \ref{sec:abel} for the  proof)
  
\barray 
& \frac{x(w) - x_0}{2 \hbar w} =  \int_{ w_0}^{ w}  dE  \, E \frac{d}{dE}
 \left( \frac{ n( 2 E)}{  E} \right) \frac{ 1}{ \sqrt{w^2 - E^2}}.
& \label{38b} 
\earray  
  
Let us apply eqs. (\ref{38}) and (\ref{38b})   to the models studied in section \ref{sec:relativity}. 

\vspace{0.5 cm}

{\bf Linear potential}

\vspace{0.5 cm}

In the case of the linear potential 

\beq
w(x) = \alpha  x, \qquad D=  (h, \infty) 
\label{39}
\eeq
one finds \cite{SL11}

\barray 
n(E)  & = &  \frac{E}{2 \pi \hbar \alpha} \left( \cosh^{-1}  \frac{E}{ 2 w_0} -
 \sqrt{ 1 - \left(  \frac{2 w_0}{ E}   \right)^2}  \right)  \label{40} \\
&   \stackrel{E>>1}{\rightarrow}  & 
\frac{E}{2 \pi \hbar \alpha} \left( \log \frac{E}{ w_0} -1 \right) + O(E^{-1})  \quad (E>> w_0) 
\nonumber 
\earray
where $w_0 = w(h) = \alpha h$. The leading term,  $O(E \log E)$, depends only on $\alpha$, and the next
to leading term, $O(E)$,  depends on $\alpha$ and $w_0$. Let us compare eq.(\ref{40}), with  
the Riemann-Mangoldt formula up to a height $t$ \cite{E74}

\beq
\langle n(t) \rangle \simeq \frac{t}{2 \pi} \left( \log \frac{ t}{2 \pi } -1 \right) + \frac{7}{8} + O(t^{-1}), \qquad t >> 1. 
\label{41}
\eeq
The first two leading terms in this  formula agree with those of (\ref{40})
under the identifications

\beq
t = \frac{ E}{ \hbar \alpha}, \qquad h = 2 \pi \hbar.
\label{42}
\eeq
Setting  $\alpha$ to 1, we recover  the model I. 
Notice that   the constant term in (\ref{40}) vanishes,  unlike in  Riemann's formula
where it is given by $7/8$.

\vspace{0.5 cm}

{\bf Linear -log model}

\vspace{0.5 cm}

One may ask which modification of $w(x)$ would yield    a counting function
$n(E)$ similar to  (\ref{40}),  but   containing   a constant term, i.e.

\beq
n(E) =  \frac{E}{2 \pi \hbar} \left( \cosh^{-1}  \frac{E}{2  w_0} - \sqrt{ 1 - \left( \frac{ 2 w_0 }{ E} \right)^2} \right) + \mu.
\label{58-6} 
\eeq
 To  find this potential we use eq.(\ref{38b}) 
 
 \barray 
 x- x_{0} &  = &  w - w_0 +  \mu \hbar \log  \frac{ 1 - \sqrt{ 1 - (w_0/w)^2}} {  1 + \sqrt{ 1 - (w_0/w)^2}}
 \label{58-7} \\
 & \simeq &  w - w_0 +  2 \mu \hbar \log \frac{ w_0}{ 2 w}, \qquad w >> w_0,  \nonumber  
 \earray 
whose inverse is

\beq
w(x) \sim x + 2 \mu \hbar \log x, \qquad  x >> 1. 
\label{58-8}
\eeq
Based on this result 
we  expect that a function $w(x)$ reproducing the Riemann zeros would contain
a $\log x$ in   its asymptotic  expansion.

\vspace{0.5 cm}

{\bf The Berry-Keating model }

\vspace{0.5 cm}

The semiclassical spectrum associated to the function $w(x) = x+ h^2/x$, defined in the halfline
$D = (0, \infty)$, is given by \cite{BK11}

\barray 
n(E) & = &  \frac{ E}{ 2 \pi \hbar} \left[ K \left( 1 - \frac{ 16 h^2}{E^2} \right) - E \left( 1 - \frac{ 16 h^2}{E^2} \right) \right]
\label{58-9} \\
&\stackrel{E>>1}{\rightarrow} &  \frac{ E}{ 2 \pi \hbar}  \left( \log \frac{ E}{h} -1 \right) - \frac{ 2 h^2}{ \pi \hbar} \frac{ \log E}{E} + \dots, 
\nonumber
\earray 
where $K$ and $E$ are the elliptic integrals

\beq
K(k^2) = \int_0^{\pi/2} d x \left( 1- k^2 \sin^2 x \right)^{-1/2}, \quad
E(k^2) = \int_0^{\pi/2} d x \left( 1- k^2 \sin^2 x \right)^{1/2}.
\nonumber 
\eeq
Choosing $t = E/\hbar$ and $h = 2 \pi \hbar$, as in (\ref{42}), one gets   an agreement
with the first two terms of the Riemann-Mangoldt formula (\ref{41}),  in the large $E$ limit. 
The constant term is absent in (\ref{58-9})  which we believe is due to 
 the absence of the term $\log x$  in $w(x)$,  for large values of $x$.
All these features are shared with the linear potential analyzed previously.

\vspace{0.5 cm}

{\bf Spacetime with constant negative   curvature}

\vspace{0.5 cm}

The semiclassical spectrum associated to the  function $w(x)$ given by (\ref{g27}) and a  
 domain $D=(0, \infty)$,  is

\beq
E_n  = \hbar \omega  ( n + \mu), \qquad n(E) = \frac{E}{ \hbar \omega} - \mu, 
\label{58-1}
\eeq
where $\mu$ and  $\omega$ are related to the parameters $w_0$ and the scalar
curvature $R = - |R|$ as 

\beq
R = - \frac{1}{2 (\mu \hbar)^2}, \qquad  
\omega = \frac{ 2 w_0 }{ \mu \hbar}. 
\label{58-5}
\eeq
It is an interesting fact   that the harmonic oscillator spectrum is associated with a spacetime
with  constant negative curvature, at least semiclassically. This result does  not contradict
the  chaotic spectrum associated to models defined in 
 two dimensional euclidean spaces with constant negative curvature, since they have different spatial dimensions, i.v.
 one versus two
\cite{S42}.

\vspace{0.2 cm}

{\bf Power like models}

\vspace{0.5 cm}

To study the  semiclassical spectrum of the models defined in (\ref{g29b}), we shall
distinguish the cases $\alpha <1$ and $\alpha >1$, which have very different properties. 
For  functions $w(x)$  that grow slower than $x$,  one obtains 

\beq
0 <  \alpha < 1 \Longrightarrow n(E) \propto E^{\frac{1}{\alpha}}, \qquad E_n \propto n^\alpha 
\label{43}
\eeq
The density of states $dn/dE$ behaves as $E^{\frac{1}{\alpha} -1}$ and  diverges in the limit $\alpha \rightarrow 0$, where
it becomes  a continuum.  The model $\alpha = 0$ corresponds to a constant  $w(x)$ and it has
a continuum spectrum (see \ref{sec:abel}).   
For potentials that grow faster than $x$, one finds

\beq
  \alpha > 1  \Longrightarrow n(E) \propto  E + O( E^{\frac{1}{\alpha}}), \qquad E_n \propto n + O( n^{ \frac{1}{\alpha}})  
\label{44}
\eeq
which is a linear spectrum with corrections. In the limit $\alpha \rightarrow \infty$, one recovers
the harmonic oscillator spectrum, which is consistent with the exponential growth of $w(x)$  (recall eq.(\ref{g27})). 

\vspace{0.5 cm}

In table 1, we collect the results obtained in this section and a comparison with the 
curvature of the associated  spacetime models.  From these examples, 
we draw the following conclusions:

\begin{itemize}

\item In models where $w(x)$ becomes linear for $x >>1$,  the first two  leading terms of $n(E)$
are of order $E \log E$ and $E$. The converse is also true, i.e. the latter asymptotic behaviour
of $n(E)$ forces  $w(x)$ to be linear for $x >>1$.  Correspondingly,  the scalar curvature $R(x)$
vanishes  faster than $x^{-2}$.

\item The appearance of a constant,  in the next leading correction to  $n(E)$, requires
a logaritmic term  in $w(x)$, in addition to the linear term. 

\item Based on these arguments,  one is lead to conjecture that the function
$w(x)$, whose quantum spectrum yields the exact Riemann zeros,  is  of the form

\beq
w(x) = x + \mu  \log x  +w_{\rm fl}(x),  
\label{41b}
\eeq
where $w_{\rm fl}(x)$ represents the fluctuation part of $w(x)$. The role of $w_{\rm fl}(x)$
is  to provide the fluctuation term $n_{\rm fl}(t)$  of the Riemann-Mangoldt formula for the exact position
of the Riemann zeros,

\barray
n_R(t) & = &  \langle n(t) \rangle + n_{\rm fl}(t), \qquad n_{\rm fl}(t)  = \frac{1}{\pi} \arg \zeta \left( \frac{1}{2} + i t \right),  
  \label{41c} 
\nonumber
\earray 
where $\zeta(s)$ is the Riemann zeta function. It is not clear at the moment how to construct $w_{\rm fl}(x)$,
or even if  it  exists.  

\end{itemize}

\vspace{1 cm}

\begin{center}
\begin{small}
\begin{tabular}{|c|c|c|c|}
\hline 
Model & $w(x)$&   $n(E)$ & $R(x)$    \\
\hline
\hline
Linear potential  & $x $ & $ \frac{E}{2 \pi} \log \frac{E}{2 \pi  e} + O(E^{-1})$ & $0$  \\ 
\hline
Berry-Keating  potential  & $x+ h^2  x^{-1} $ & $ \frac{E}{2 \pi} \log \frac{E}{2 \pi  e} + O(\log E/E)$ & $-  4 h^2  \,  x^{-4}$  \\ 
\hline
Linear-log potential  & $x + 2  \mu  \log x$  & $\frac{E}{2 \pi} \log \frac{E}{2 \pi  e} + \mu + O(E^{-1})$    & $ 4  \mu  x^{-3}$     \\
\hline
Harmonic oscillator  & $w_0 \cosh  (x/2 \mu) $ & $\mu ( E/2 w_0  - 1)$ & $- 1/2 \mu^{2}$   \\ 
\hline
Sublinear  potentials  & $x^\alpha \; (0< \alpha < 1)  $ & $ E^{1/\alpha}$ & $- 2 \alpha (\alpha-1) x^{-2} > 0$  \\ 
\hline
Superlinear potentials  & $x^\alpha \; (\alpha > 1)  $ & $ E + O(E^{1/\alpha})$ & $- 2 \alpha  (\alpha-1) x^{-2} < 0$  \\ 
\hline 
\hline 
\end{tabular}
\end{small}

\vspace{0.2 cm}

\end{center}
Table 1.- Semiclassical counting function $n(E)$,  and scalar curvature $R(x)$, for the models discussed in the text. 
 We show the asymptotic behaviour of the functions $n(E)$ and $R(x)$, i.e. $E>>1, x>>1$, with $\hbar =1$.

\section{ Quantization}
\label{sec:quantization} 

To quantize  the classical Hamiltonian  (\ref{1}) we shall choose the normal ordering prescription

\beq
\hat{H} = u(x) \, \hat{p} \, u(x) + v(x) \, \frac{1}{ \hat{p}} \, v(x), \qquad x  \in D= ( \ell_x, \infty) 
\label{h1}
\eeq
where $\hat{p} = - i \hbar d/dx$ and $1/\hat{p}$ is the one dimensional Green function

\beq
\langle x | \, \frac{1}{ \hat{p} } \,  |y \rangle = - \frac{i}{\hbar} \, \theta(y-x), 
\label{h2}
\eeq
written in terms of  the Heaviside step function. 
This normal ordering generalizes the one   in references \cite{SL11,BK11}.  
One can check that $1/\hat{p}$ is the inverse of $\hat{p}$ acting on wavefunctions $\psi(x)$ that vanish in the limit $x \rightarrow \infty$, 

\barray
\left( \hat{p} \;  \frac{1}{ \hat{p}} \;  \psi \right)(x) & = &  - \frac{\partial}{\partial x} \int_{\ell_x}^\infty dy \, \theta(y-x) \, \psi(y) = 
\int_{\ell_x}^\infty dy \, \delta(y-x) \, \psi(y) = \psi(x),  \nonumber   \\
\left(   \frac{1}{ \hat{p}} \; \hat{p} \; \psi \right)(x) & = &  -  \int_{\ell_x}^\infty dy \, \theta(y-x) \,\frac{\partial \psi(y) }{\partial y}  = 
- \lim_{y \rightarrow \infty} \psi(y)  + \psi(x),  \label{h3}
\earray 
where we have assume that $x > \ell_x$.  
There exist   other possible normal
orderings  defining $\hat{H}$, as for example 
$\frac{1}{2} ( u^2(x) \,  \hat{p} + \hat{p} \,  u^2(x) + v^2(x) \,  \hat{p}^{-1}  + \hat{p}^{-1} \, v^2(x) )$
(see also \cite{BK11}). 
However, the choice (\ref{h1}) yields a  Schroedinger equation which, as we shall see below,   is equivalent to a second order differential
equation, suplemented with a non local 
boundary condition. Our construction 
parallels in that way   the Schroedinger equation arising from the quantum
Hamiltonian $\hat{H} = \hat{p}^2/2m  + V(x)$. The action of (\ref{h1}) on a wavefunction $\psi(x)$ is   given by

\barray 
( \hat{H}  \psi)(x) &  = &  - i  \hbar \,  u(x)  \frac{d}{dx}  \left\{ u(x)   \psi(x) \right\}   \label{h4} \\
&  & -  i  \hbar^{-1}      \int_{\ell_x}^\infty dy  \, v(x)  \;   \theta(y-x)  \, v(y)   \psi(y).  
\nonumber 
\earray
This operator will have a real spectrum if it is self-adjoint, which requires, first of all, that it be symmetric \cite{GP90}

\beq
\langle \psi_1 |  \hat{H}   \psi_2 \rangle  =  \langle \hat{H}  \psi_1 | \psi_2 \rangle , \qquad \forall \psi_1, \psi_2 \in {\cal D} (\hat{H}), 
\label{h5} 
\eeq
where ${\cal D} (\hat{H})$ is the domain of the operator ${\hat H}$. To study this condition,  
 let us define the quantity \cite{AIM05}

\barray 
\langle \psi_1 |  \hat{H}   \psi_2 \rangle - \langle \hat{H} \psi_1 | \psi_2 \rangle & = & - i  \, \Omega_{12}.  
\label{h6} 
\earray 
Using (\ref{h4}) one finds

\barray 
\Omega_{12} & = & 
\hbar  \int_{\ell_x}^\infty  dx   \frac{d}{dx} ( u^2(x)   \, \psi_1^*(x) \psi_2(x) )  \label{h7} \\
& + &  \hbar^{-1}  
 \int_{\ell_x}^\infty \int_{\ell_x} ^\infty  dx \, dy \, v(x) v(y) \psi_1^*(x) \psi_2(y)    (  \theta(x-y)+  \theta(y-x)) 
\nonumber \\
& = &  - \hbar  u^2 (\ell_x)\psi_1^*(\ell_x) \psi_2(\ell_x) +  \hbar^{-1}   \int_{\ell_x}^\infty \int_{\ell_x}^\infty  dx \, dy \, v(x) v(y)  \psi_1^*(x) \psi_2(y), 
\nonumber 
\earray
where we have assumed  that   $\lim_{ x \rightarrow \infty} u(x)  \,  \psi_{1,2}(x) =0$. 
Hence  $\hat{H}$  is a    symmetric operator  iff  $\Omega_{12} = 0$ which, in view of eq.(\ref{h7}),
is guaranteed if $\psi_{1,2}$ satisfy the equation

\beq
e^{ i \vartheta}\,  \hbar \,  u(\ell_x)  \,  \psi(\ell_x) +    \hbar^{-1}   \int_{\ell_x}^\infty  dx \,  v(x)  \psi(x) = 0, 
\label{h8}
\eeq
where $\vartheta$  parameterizes the selfadjoint extensions  of $\hat{H}$. 
  The Schroedinger equation of  the Hamiltonian (\ref{h4}) is given  by

\beq
- i  \hbar \, u(x)  \frac{d}{dx}  \left\{ u(x)   \psi(x) \right\}  - i   \hbar^{-1}   \int_{\ell_x}^\infty dy  \, v(x)  \;   \theta(y-x)  \, v(y)   \psi(y)   = E \, \psi(x), 
\label{h9}
\eeq
which we write as

\beq
  \hbar \, \frac{ u(x) }{ v(x)}  \frac{d}{dx}  \left\{ u(x)   \psi(x) \right\} - \frac{ i E}{v(x)} \psi(x)  +    \hbar^{-1}   \int_{\ell_x}^\infty dy   \;   \theta(y-x)  \, v(y)   \psi(y)   = 0. 
\label{h10}
\eeq
Taking a derivative  respect to $x$ and letting  $x= \ell_x$ one gets 

\barray 
\frac{d}{dx} \left[ \hbar \frac{u(x)}{v(x)}   \frac{d}{dx}  \left\{ u(x)   \psi(x) \right\} - 
\frac{ i E}{ v(x)} \psi(x) \right] -  \hbar^{-1} v(x) \psi(x)  &= &  0, 
\label{h11} \\
  \left[ \hbar \, \frac{ u(x) }{ v(x)}  \frac{d}{dx}  \left\{ u(x)   \psi(x) \right\} - \frac{ i E}{v(x)} \psi(x) \right]_{x=\ell_x}   +    \hbar^{-1}   \int_{\ell_x}^\infty dy    \, v(y)   \psi(y)   & = &  0. 
\label{h12}
\earray 
It would seem  that one has  to solve simultaneously eqs.(\ref{h11}) and  (\ref{h12}). 
However, for wavefunctions that decay sufficiently fast at infinity, 
 the latter equation follows from the integration of the former one, dropping a
 term at infinity.   Hence,  the Schroedinger equation (\ref{h9}) is equivalent to the second order differential 
 equation  (\ref{h11}), which 
 justifies   the normal ordering  (\ref{h1}). 

Eqs. (\ref{h11}) and (\ref{h12}) exhibit the general covariance discussed in the 
section \ref{sec:classical}.  This can be shown using the transformation laws of $u(x), v(x)$ given in section \ref{sec:classical},
together with that of $\psi(x)$, 

\beq
\psi'(x') \, (dx' )^{1/2} = \psi(x) \, (dx)^{1/2}.
\label{h16}
\eeq
which preserves   the form of the scalar product of the Hilbert space under diffeomorphism, 

\beq
\int_{\ell'_x}^\infty  dx'  \, \left( \psi'(x') \right)^*  \psi'(x') = \int_{\ell_x}^\infty dx \,  \psi^*(x) \, \psi(x). 
\label{h17}
\eeq
As was done in the classical and semiclassical analysis, it is convenient to work
in the symmetric gauge.  Let us  transform eq.(\ref{h11}) into that gauge.
First we write it as  

\barray 
\hbar \frac{u(x)}{v(x)}
\frac{d}{dx} \left[ \hbar \frac{u(x)}{v(x)}   \frac{d}{dx}  \left\{ u(x)   \psi(x) \right\} - 
\frac{ i E}{ u(x) v(x)}  u(x) \psi(x) \right] -  u(x) \psi(x)  &= &  0, 
\nonumber
\earray
and make the transformation  $x \rightarrow x'$ (\ref{13}) to the symmetric gauge (i.e $dx' = dx \,   v(x)/u(x)$),

\barray 
 \hbar
\frac{d}{dx'} \left[  \hbar  \frac{d \phi(x')}{dx'}   -  \frac{ i E}{ w(x')}  \phi(x') \right] -   \phi(x' )  &= &  0, 
\label{h14}
\earray 
where $w(x')$ is given by eq.(\ref{14}),  and  the  function $\phi$ is defined as 

\beq
\phi(x') = u(x(x')) \, \psi(x(x')).
\label{h15}
\eeq
Notice that  $\phi(x)$ is a scalar function, i.e. $\phi'(x') = \phi(x)$. Now we rename $x'$ as $z$, 
and write  eqs.  (\ref{h14}) and (\ref{h8}),  for an eigenfunction $\phi_E(z)$, with energy $E$, 
as

\barray 
\hbar \frac{d}{dz} \left(  \hbar \frac{d \phi_E(z)}{dz} - \frac{i E}{ w(z)} \phi_E(z)  \right) - \phi_E(z)  & = &  0, \quad z \geq z_0
 \label{h18} \\
\qquad \quad \hbar  \,  e^{i \vartheta}  \phi_E(z_0) +  \hbar^{-1}  \int_{z_0}^{\infty}  dz \, \phi_E(z) & = &  0.  \label{h19}
\earray
The norm of the wave function $\psi(x)$ becomes in this gauge

\beq
\langle \psi| \psi \rangle = \int_{\ell_x}^\infty dx \, \psi^*(x) \, \psi(x) = \int_{z_0}^\infty  \frac{dz}{w(z)} \, \phi^*(z) \, \phi(z).  
\label{h20}
\eeq

Let us next discuss the behaviour of the eigenfunctions  under the time reversal transformation. The classical model
has the property that if $ \left\{ x(t), p(t) \right\}$ is a trajectory with energy $E$, then  $\left\{ x(t), -p(t) \right\}$ is a trajectory
with energy $-E$. Under time reversal,  the quantum Hamiltonian (\ref{h1}) changes sign, which leads
us to expect  a relation between eigenfunctions with energies $E$ and $-E$. Indeed,  taking  the complex conjugate of
eqs.(\ref{h18},\ref{h19}), and comparing  them with those  for an eigenfunction $\phi_{-E}(z)$,
one finds 

\beq
 \phi^*_{E}(z) \propto  \phi_{-E}(z) \Longleftrightarrow e^{i \vartheta} = e^{- i \vartheta} \Longleftrightarrow \vartheta = 0 \,  \; {\rm or} \; \, \pi
\label{h23}
\eeq
Hence, if $\vartheta =0$ or $\pi$, the spectrum displays  the time reversal  symmetry $E \leftrightarrow -E$
(this fact was already observed in \cite{BK11}). The difference between
these cases resides  in the existence of a
  zero energy state.  The solution of equation (\ref{h18}) for  $E=0$ is given by

\beq
\phi_{E=0}(z)  = A \, e^{- z/\hbar} + B \,  e^{ z/\hbar}
\label{h24}
\eeq
The exponential growing term $e^{z/\hbar}$  will typically yield unnormalizable wave functions, so we
only consider the decaying  term. In this case the nonlocal boundary condition (\ref{h19}) becomes

\beq
\hbar ( e^{i \vartheta } + 1) e^{ - h/\hbar} = 0 \Longrightarrow \theta = \pi 
\label{h25}
\eeq
so,  only for  this value of $\vartheta$ is $\phi_0 = e^{-z/\hbar}$ an eigenfunction of the Hamiltonian with a norm given by

\beq
\langle \psi_{0}| \psi_{0} \rangle =   \int_{z_0}^\infty  \frac{dz}{w(z)}  \, e^{- 2 z/\hbar} 
\label{h26}
\eeq
which will be finite  for a large class of models that includes the ones studied in sections \ref{sec:relativity}
and \ref{sec:semiclassical}. 
In summary, when $\vartheta =0$, the spectrum of the Hamiltonian (\ref{h1}) contains time reversed
pairs $\{ E_n, - E_n \}$, excluding the zero energy, while if  $\vartheta = \pi$, in addition
to the time reversed pairs, there is  a normalizable zero energy state. Depending on the form
of $w(z)$, the spectrum may, or may not,  contain a continuum part. 

\subsection{The model I}

In the following subsection we shall  apply the previous formalism 
to   the model with a linear potential $w(x)$  \cite{SL11}

\barray 
w(z)  & = &   z, \qquad   D = ( z_0, \infty).  
 \label{m1} 
\earray
 The eigenfunctions of the Hamiltonian (\ref{h1})
are the solutions of  the differential equation (\ref{h18}), which in this case read ($\hbar =1$)

\beq
\phi''(z) - \frac{i E}{ z} \phi'(z) + \left( \frac{i E}{ z^2 } -1 \right)  \phi(z)  = 0.
\label{m2bis}
\eeq
The solutions of this equation  are given essentially by  the modified Bessel functions, 
but only the $K$-Bessel function gives a normalizable solution 

\beq
 \phi_\nu(z)  =  A_\nu  \, z^{ 1 - \nu}   \, K_{\nu}(z),   \qquad \nu = \frac{1}{2} - \frac{ i E}{2}, 
\label{m3}
\eeq
where $A_\nu$ is the normalization constant. Using eq.(\ref{h15}) one obtains 

\beq
\psi_E(z) = A_\nu  \,  z^{ \frac{i E}{2}}  \, K_{\frac{1}{2}- \frac{i E}{2}}(z), \quad z \geq z_0, 
\label{m4}
\eeq
whose  asymptotic behaviour  is 

\beq
\psi_E(z) \propto 
\left\{
\begin{array}{cc}
z^{- \frac{1}{2} + i E},  & z_0 < z << E/2 \\
z^{- \frac{1}{2} +  \frac{i E}{2}} e^{- z},  & z >> E/2. \\
\end{array}
\right.
\label{m5}
\eeq
which shows that in the region $z_0 < z << E/2$,  the function $\psi_E$ is given
approximately by the eigenfunction of the Hamiltonian $\hat{H} = (x \hat{p}+ \hat{p} x)/2$ \cite{BK99,S07a,TM07}. 
Notice that  $x_M = E/2$ coincides with the maximal elongation of the classical particle. Beyond this value
the wave function decays exponentially, as corresponds to the particle entering the classical forbidden region
$x > x_{\rm M}$. The equation for the eigenenergies can be  obtained plugging (\ref{m3}) into (\ref{h19}),  and 
using the integral \cite{GR} 

\beq
\int_{z_0}^\infty dz \, z^{1- \nu} \, K_{\nu}(z) = z_0^{1-\nu} \, K_{\nu -1}(z_0),
\label{m6}
\eeq
with the result  \cite{SL11} 

\beq
 e^{ i \vartheta} K_{\nu} ( z_0) + K_{\nu -1}(z_0) = 0 \rightarrow  e^{ i \vartheta} K_{\frac{1}{2} - \frac{ i E}{ 2 }} ( z_0) +
K_{\frac{1}{2} + \frac{ i E}{ 2}} ( z_0) = 0.
\label{m23}
\eeq
The asymptotic behaviour of the $K$-Bessel function

\barray 
K_{\frac{1}{2}  + 
\frac{ i t}{2}}(z)  & \sim &  \sqrt{  \frac{\pi}{z} }   \, e^{ - \pi t/4}    \left(
\frac{t}{ z e} \right)^{ i t/2} , \quad t >> 1 
\label{b11} 
\earray 
yields the asymptotic limit of (\ref{m23})

\beq
\cos \left( \frac{E}{ 2 } \log \left( \frac{E}{   z_0  e} \right) - \frac{ \vartheta}{2} \right) = 0, \qquad E>> 1,  
\label{m24}
\eeq
so that  the eigenenergies  $E >0$ behave  as 

\beq
n(E) = \frac{E}{ 2  \pi } \left(  \log  \frac{E}{  z_0} - 1 \right)  - \frac{ \vartheta}{2 \pi} - \frac{1}{2} \in {\cal Z}  
\label{m25} 
\eeq
The two leading terms  in this equation  agree with the Riemann formula (\ref{41}) 
and the semiclassical result (\ref{40}), with the identification $z_0= 2 \pi$,
which is the same as in eq.(\ref{42})  (notice that $w_0 =z_0=h$). 
We must set $\vartheta= 0$ to guarantee  time reversed eigenenergies
in analogy with the symmetry of the Riemann zeros on  the real axis. 
The quantization of the model  brings in a constant factor $-1/2$ in the counting formula (\ref{m25}), 
so that the factor $7/8$ of Riemann's formula remains unexplained. In this respect,  
we recall the comment made in the previous section that adding a term $\log x$ to $w(x)$, 
can give rise to this constant term in the counting formula.

\section{Conclusions}

In this paper we have studied the properties of a  family 
of one dimensional classical models, and their quantized version,   that  are extensions of the well known 
$xp$ model. A fundamental property of these models is that they are covariant 
under general coordinate transformations, so that they can be  organized into equivalent classes
that describe the same Physics. This fact suggests some 
 some sort of universality  that  
could perhaps be  explained using renormalization group arguments,  as the ones employed
in reference \cite{S05}, where the operator $1/\hat{p}$ was also considered. 

 General covariance manifests itself most 
clearly in the Lagrangian formulation of  a relativistic particle moving  in a
1+1 dimensional spacetime. The geometrical properties of this spacetime turn out to be related 
to the spectral properties of the associated 
quantum model in an deep way. To wit, when the curvature of spacetime vanishes fast enough at infinity, one obtains a spectrum 
that coincides with the Riemann zeros on average. It is tempting to think that certain  fluctuations of 
asymptotically  flat metrics, may 
cause fluctuations in the spectrum so as to accurately reproduce the  Riemann zeros. 
It is expected that these fluctuations are determined by the prime numbers, but
the precise manner in which this might occur is unclear. This general class of models
can arise as effective  descriptions of the dynamics of an electron moving in the plane
under the action of a perpendicular uniform magnetic field and 
 a electrostatic potential of the form $V(x,y) = U(x) y + V(x)/y$, where
$x$ and $y$ are the two dimensional coordinates \cite{ST08}.  Further investigation will be needed
to clarify these issues.

We hope the results presented here shed new light 
on the spectral interpretation of the Riemann zeros, 
stimulating the research into  such a interdisciplinary topic

\vspace{1cm} 

{\bf Acknowledgements.- }  I am   grateful to Manuel Asorey, Michael  Berry, 
Jon  Keating,  Jeff Lagarias,   Javier  Rodr\'{\i}guez-Laguna, Giuseppe Mussardo  and Paul  Townsend, for fruitful discussions.  
I also thanks  Hubert  Saleur and Jesper Jacobsen for the invitation to
participate in  the Program "Advanced Conformal Field Theory and Applications"
in the Centre \'{E}mile Borel at IHP (Paris), where this work was completed. 
This work has  been financed  by the Ministerio de  Ciencia e Innovaci\'on, 
Spain (grant FIS2009-11654) and  Comunidad de Madrid (grant QUITEMAD).

\appendix

\section{Semiclassical Abel like  inversion formula}
\label{sec:abel} 

The aim of this appendix is to derive a formula that, under certain conditions, 
permits to construct  the potential $w(x)$ that reproduces a given semiclassical counting
formula $n(E)$. For Hamiltonians of the form
$H = p^2/2m + V(x)$,  the answer can be obtained using the  Abel inversion
method  \cite{GP90}. We shall  also use Abel's  method for Hamiltonians of the form $H = w(x) (p + 1/p)$.
First of all,  we shall assume that $w(x)$ is a monotonic increasing function in the domain
$D = (x_0, \infty)$, so that it is invertible, $x = x(w)$. 
In this case the semiclassical formula (\ref{38}) becomes 

\beq
n(E) = \frac{1}{2 \pi \hbar} \int_{x_0}^{x_M} \frac{dx}{ w(x)} \sqrt{ E^2 - 4 w^2(x)}, \qquad 
  E = 2 w(x_M), 
\label{ab1}
\eeq
which  we write  as 

\beq
   \frac{2 \pi \hbar \, n(E)}{E}  =  \int_{x_0}^{x_M}  dx  \, \sqrt{ \frac{1}{w^2(x)}  - \left( \frac{2}{ E} \right)^2  }.  
\label{51}
\eeq
Making  the change of variables (with $E>0$) 

\beq
r(x) = \frac{1}{w(x)}, \quad s = \frac{2}{E}, \quad r_0 = \frac{1}{w_0}  \geq  s , \; w_0 = w(x_0),   
\label{53}
\eeq
and using the inverse $x(w)$, we transform  (\ref{51}) into

\beq
f(s) \equiv \pi \hbar \, s \,  n(s)    =  \int_{x_0}^{x_M}  dx  \, \sqrt{ r^2(x)  -s^2   } =  \int^{s}_{r_0}  dr \,   \frac{d x}{dr}  \, \sqrt{ r^2  -s^2   }. 
\label{54}
\eeq
Differentiating  with  respect to $s$ yields

\beq
\frac{d f(s)}{ds}    =    \int_{s}^{r_0}  dr  \,  \frac{d x}{dr}  \,  \frac{  s}{ \sqrt{ r^2  -s^2   }}. 
\label{55}
\eeq
Next,  we  use   the  chain of identities

\barray 
& \int_y^{r_0} ds \,  \frac{ 1}{ \sqrt{ s^2 - y^2  }} \,  \frac{d f(s)}{ds}     =   \int_y^{r_0} ds \,   \int_{s}^{r_0}  dr  \,  \frac{d x}{dr}
   \,  \frac{  s}{ \sqrt{( s^2 - y^2)( r^2  -s^2  ) }} & \label{56}  \\
& =    \int_y^{r_0} dr \,   \frac{d x}{dr}  \int_{y}^{r}  ds  
   \,  \frac{  s}{ \sqrt{( s^2 - y^2)( r^2  -s^2  ) }}   =  \frac{ \pi}{2}   \int_y^{r_0} dr \,   \frac{d x}{dr}  = 
   \frac{ \pi}{2}  ( x_0 -  x(y)), &    \nonumber  
\earray
that  can be   obtained from   the Frobenius theorem and the integral

\beq
 \int_a^b dx \frac{ x}{ \sqrt{ ( x^2 - a^2)(b^2 - x^2)}} = \frac{ \pi}{2} , \qquad 0 < a < b.  
 \label{57}
 \eeq
Replacing $f(s)= \pi \hbar s n(s)$ into (\ref{56}), and undoing the change of variables (\ref{53}),  yields
finally 

\barray 
& \frac{x(w) - x_0}{2 \hbar w} =  \int_{ w_0}^{ w}  dE  \, E \frac{d}{dE} \left( \frac{ n( 2 E)}{  E} \right) \frac{ 1}{ \sqrt{w^2 - E^2}}.
\label{58}
\earray 
It is not guaranteed a priori
that $x(w)$ is an invertible function. This  fact  may   restrict  the   functions $n(E)$  that can be obtained 
as semiclassical spectrum. 
Another   issue concerning (\ref{58}) is the following. 
Adding to   $n(E)$  a term linear in $E$,
produces the same function $w(x)$, as can be easily seen from (\ref{58}). We  lost track  of this linear term 
when taking the derivative respect to $s$ in eq.(\ref{54}). To recover this term, one has to plugg the function $w(x)$, 
obtained from (\ref{58}), back into (\ref{ab1}), and read  the linear term in $E$.

It is worth to  compare these semiclassical formulas  with those associated to the standard  
Hamiltonian $H = p^2 + V(x)$. 
The analogue of eq. (\ref{ab1}),  for  an even potential $V(x)$,  is given by

\beq
n(E) =   \frac{2}{ \pi \hbar}  \int_{x_0}^{x_m}   dx  \, \sqrt{ E - V(x) }, \qquad |E| = V(x_m), 
\label{se45}
\eeq
where we have ignored the Maslow phase,  and the analogue of (\ref{58}) is given by 

\beq
x(V) = \hbar \int_{V_0}^V  d E \;  \frac{ d n(E)}{ d E} \frac{1}{ \sqrt{ V - E}}. 
\label{se47}
\eeq
The latter equation was used by Wu and Sprung to obtain a potential $V(x)$ whose
semiclassical spectrum coincides in average with the Riemann zeros (\cite{WS93}, for a review see \cite{SH11}),

\beq
x(V)  = \frac{ \sqrt{V}}{\pi} \log \left( \frac{ 2 V}{ \pi e^2} \right)  \Longrightarrow V(x) \propto \left( \frac{ x}{ \log x} \right)^2, 
\qquad x, V  >> 1 
\label{se48}
\eeq
 This result  was used as a seed for a  numerical reconstruction of a potential 
whose spectrum matches a large number of Riemann zeros lying at the bottom part of the critical line. Quite interestingly,  that  potential 
has a fractal structure whose  dimension is nearby  the value  $1.5$.  A problem with this approach is that
the Hamiltonian is  time reversal invariant, a fact which does not agree with the  distribution of the 
 Riemann zeros which follow the GUE statistic which is characteristic of time reversal breaking random Hamiltonians
 (see \cite{SH11}) for a discussion on this issue). 

Independently of the previous works, Mussardo employed (\ref{se47}) to find a potential whose spectrum 
behaves, in average,  like the prime numbers \cite{M97}.  The prime number theorem (PNT) states that the number of primes
up to $x$ behaves asymptotically as $\pi(x) \sim x/\log x$, so that their   density decreases as $d \pi(x)/dx  \sim 1/\log x$,
as conjectured long ago by Gauss and Legendre \cite{E74}.  The PNT implies that the $n^{\rm th}$-prime number $p_n$ grows 
roughly as $p_n \sim n \log n$.  Mussardo noticed that this growth allows one to find a quantum mechanical  model
whose energies are the primes numbers. Choosing  the leading term in the expansion of the Riemann formula for $\pi(x)$,

\beq
\pi(x) \sim Li(x) = \int_2^x \frac{dy}{\log y} 
\label{49}
\eeq
he  obtained  

\beq
x(V)  \sim \frac{ \sqrt{V}}{ \log V}  \Longrightarrow V(x) \sim ( x \log x)^2,  
\quad (x, V>>1) 
\label{50}
\eeq
This result  was also used as a seed for finding a potential whose spectrum matches
precisely the lowest prime numbers \cite{SH11}. 
As in the case of the Riemann zeros, the prime potential  has a fractal structure
with dimension  near  2.  The difference  in fractal dimensions, $1.5$ versus $2$,  is consistent with the fact that the Riemann zeros
are less random that the primes numbers. The former ones satisfy  the  GUE statistics and the latter follow  an almost  Poissonian  statistics.  
It is rather remarkable the {\em proximity}  of the  prime number/Riemann zeros potentials to the harmonic oscillator potential. 
In table 2 we summarize these semiclassical results together with other well known  cases. 
 
\vspace{0.5 cm} 

\begin{center}
\begin{small}
\begin{tabular}{|c|c|c|c|c|}
\hline 
Model & $V(x)$& $x(V)$  & $ E_n$   &  $n(E)$   \\
\hline
\hline
Free particle   & cte  & $ 0 \leq |x|  \leq \infty$ & continuum  & -   \\ 
\hline
Harmonic oscillator  & $x^2 $ & $\sqrt{V}$ & $n$ & $E$  \\ 
\hline
Potential well  & cte & $ 0 \leq |x| < 1  $  & $ n^2$   &  $ \sqrt{E}$   \\
\hline
Riemann zeros  & $( \frac{x}{ \log x} )^2 $ & $\sqrt{V} \log V$ & $  \frac{2 \pi n}{\log n}$ & $\frac{E}{2 \pi} \log \frac{E}{2 \pi e} $  \\ 
\hline
Prime numbers   & $( x \log x)^2 $  & $ \frac{ \sqrt{V}}{ \log V}  $  & $ n \log n $   &  $ \frac{E}{ \log E}$   \\
\hline 
\end{tabular}
\end{small}

\vspace{0.2 cm}
Table 2.- Semiclassical spectrum associated to the classical Hamiltonian $H = p^2 + V(x)$. 

\end{center}

\section{Quantization of   $H = \hat{p}  + \ell_p^2/ \hat{p}$}
\label{sec:kondo} 

In standard textbooks of Quantum Mechanics it is taught  that the momentum operator $\hat{p} = - i \hbar d/dx$
is self-adjoint acting in the Hilbert space of square integrable functions,  $L^2(D)$, in two  cases: i) 
 $D = \Rmath$ is  the  real line,  and ii)  $D = (a, b)$ is a finite interval of the real line \cite{GP90,BFV01}. 
In case i),  the operator $\hat{p}$  is essentially self-adjoint, and in case ii) $\hat{p}$  admits 
infinitely many self-adjoint extensions characterized by 
  the boundary condition $\psi(b) = e^{i \vartheta} \psi(a)$, where $\vartheta \in [0, 2 \pi)$. 
However, $\hat{p}$ is not self adjoint 
when $D$ is the  halfline,  $D=( 0,  \infty)$, and  therefore its spectrum is not real.
A {\em solution}  of this problem is suggested by the model we have discussed in this paper. Indeed, let us choose 
the simplest non vanishing potential $w(x)$, namely a constant

\beq
w(x) = \ell_p  \Longrightarrow \hat{H} = \hat{p} + \frac{\ell_p^2}{\hat{p}}, 
\label{ap1}
\eeq
defined on the halfline $D=(0, \infty)$. This model is equivalent to (\ref{g264}), via 
a scale transformation which gives the relation $c = \ell_p$. 
We shall show below that
$\hat{H}$ is self adjoint and that its spectrum is a  continuum and eventually  a bound state. 
This model illustrates in a simple example the more complicated models considered
in the main body of the paper. In  spite of its  simplicity,  this model shares
several  features  with the so called Kondo model in Condensed Matter Physics, 
which suggests  that it may  have other  applications.

To quantize (\ref{ap1}),  we  follow   the steps
of section \ref{sec:quantization}. The Schroedinger equation is given by 

\beq
- i \hbar  \frac{d \psi(x)}{dx} - i  \frac{\ell_p^2}{\hbar} \int_{0}^\infty dy \, \vartheta(y - x) \psi(y)  = E \psi(x).
\label{ap2}
\eeq 
Taking one derivative respect to $x$, gives 

\beq
-  \hbar^2   \frac{d^2 \psi(x)}{dx^2} +  i  E \hbar \frac{d \psi(x)}{dx}  + \ell_p^2 \psi(x) = 0, 
\label{ap3}
\eeq 
whose general solution is

\beq
\psi_E(x) = A(E) \,   e^{i k_+(E)  x} + B(E)  \,  e^{i k_-(E)   x}, 
\label{ap4}
\eeq
where

\beq
k_\pm(E) = \frac{1}{2 \hbar}  \left( E \pm \sign  (E)  \sqrt{ E^2 - 4 \ell_p^2} \right). 
\label{ap5}
\eeq
We are assuming in (\ref{ap5}) that $E$ is real, for other values we  replace $\sign(E)$ by $\pm E/|E|$. 
Using (\ref{ap5})  one can compute the von Neumann deffect indices $n_\pm$, which give the number of linearly
independent solutions of the equation 

\beq
n_\pm = {\rm dim} \;  \{\psi_\pm \, \, |   \hat{H} \, \psi_\pm  = \pm i z    \, \psi_\pm,   \; \; {\rm Im} \, z >0  \}. 
\label{ap6}
\eeq
Choosing $z = 2 \ell_p c \; (c >0)$, we get two normalizable solutions of (\ref{ap5}), namely

\barray
\psi_+(x)  &  = &  A \,  {\rm exp} \left[  - \frac{\ell_p \, x }{\hbar} \left( c + \sqrt{c^2 + 1} \right)  \right]  \Longrightarrow n_+ = 1, \label{ap7} \\
\psi_-(x)  &  = &  A  {\rm exp} \left[  - \frac{\ell_p \, x }{\hbar} \left(- c + \sqrt{c^2 + 1} \right) \right]   \Longrightarrow n_- = 1,
\nonumber
\earray
hence  $n_+ = n_-=1$, which implies, by  the von Neumann theorem, 
that the operator $\hat{H}$ is self-adjoint, with an  infinitely many extensions parametrized by the group $U(1)$ \cite{GP90,AIM05}.
The spectrum of $\hat{H}$ is given by  two intervals whose boundaries are $\pm 2 \ell_p$ and $\pm \infty$. 
and a bound state with eigenvalue $E_0$

\beq
{\rm spec} \; \hat{H} = {\cal C} \cup \{E_0 \} = 
 (- \infty, - 2 \ell_p) \cup (2 \ell_p, \infty) \cup  \{E_0 \}.
\label{ap8}
\eeq
We denote by ${\cal C}$  the continuum part of the spectrum. 
It is convenient to parametrize the two branches of the continuum as follows

\beq
E = 2 \ell_p \eta \cosh u, \qquad \eta =  \sign(E) = \pm 1,  \qquad u > 0, 
\label{ap9}
\eeq
in which case the momenta (\ref{ap5}) become

\beq
k_+(E) = \frac{ \eta \,  \ell_p}{ \hbar} e^u, \qquad k_-(E) = \frac{ \eta  \, \ell_p}{ \hbar} e^{-u}, 
\label{ap10}
\eeq
and the wave function (\ref{ap4})

\beq
\psi_E(x) = A(E) \,   e^{ i \eta \ell_p   x e^u/\hbar} + B(E)  \,  e^{i  \eta \ell_p  e^{-u} /\hbar}. 
\label{ap11}
\eeq
The discrete eigenvalue of $\hat{H}$ appears for 
 $|E_0| < 2 \ell_p$ with a normalizable  eigenfunction corresponding 
to the momenta $k_+$, i.e.

\beq
|E_0| < 2 \ell_p \Longrightarrow \psi_{E_0} (x)  = C \, e^{- k_0 x}, \quad k_0 = - i k_+ = \frac{1}{2 \hbar} 
\left( - i E_0  + \sqrt{ 4 \ell_p^2 - E_0^2} \right).  
\label{ap12}
\eeq
To find the value $E_0$,  we impose the non local boundary condition  (\ref{h8}), which guarantees that $\hat{H}$
is an hermitean operator,

\beq
- e^{i \vartheta } \psi_{E_0}(0) + \frac{\ell_p}{ \hbar} \int_0^\infty dx \, \psi_{E_0}(x) =0, 
\label{ap13}
\eeq
where $\vartheta$ is the parameter that characterizes the self-adjoint extension of $\hat{H}$.  In eq.(\ref{h8}), we  made the shift
$\vartheta \rightarrow \vartheta + \pi$, so that the first term in (\ref{ap13}) changed  its sign. Plugging (\ref{ap12})
into (\ref{ap13}) one finds

\beq
k_0 = \frac{ \ell_p}{ \hbar} e^{- i \vartheta}, 
\label{ap14}
\eeq
which gives 

\beq
E_0 = 2 \ell_p \sin \vartheta, \qquad \vartheta \in \left(- \frac{ \pi}{2}, \frac{ \pi}{2} \right). 
\label{ap15}
\eeq
The restriction on  $\vartheta$ comes from the relation  $\cos \vartheta \propto {\rm Re}  \,  k_0 > 0$.  
If  $\pi/2 < \vartheta \leq \pi$, the equation (\ref{ap13}) is not satisfied and therefore there is no a bound state.
We shall restrict below to the case $|\vartheta| < \pi/2$. 
If $\vartheta =0$, one gets  $E_0=0$, so that the spectrum is time reversal symmetric.
 In the limits  $\vartheta \rightarrow  \pm \pi/2$, 
one has $E_0 =   \pm 2 \ell_p$, and $k_0= \mp i \ell_p/\hbar$, so that the 
eigenfunction (\ref{ap12}), becomes a plane wave. The constant $C$ in eq.(\ref{ap12}), is fixed by the normalization
of the wave function

\beq
\int_0^\infty dx \, |\psi_{E_0}(x) |^2 = 1 \Longrightarrow C = \sqrt{ \frac{ 2 \ell_p \cos \vartheta}{\hbar}}. 
\label{ap16}
\eeq
The measure of the size of the bound state is given   by   the average of $x$, 

\beq
\langle x \rangle = \int_0^\infty dx \, x  |\psi_{E_0}(x) |^2 = \frac{ \hbar}{ 2 \ell_p \cos \vartheta},   
\label{ap17}
\eeq
and  diverges in the limit  $\vartheta \rightarrow \pm \pi/2$. The operator $\hat{H}$ is self adjoint, then the spectral
theorem implies that its eigenfunctions $\psi_E$ form an orthornormal basis, namely

\barray 
\langle  \psi_{E_0}  | \psi_{E_0} \rangle  & = &  \int_0^\infty dx \, \psi_{E_0} (x) \, \psi_{E_0}(x) = 1,  
\label{ap18}  \\
\langle  \psi_{E_0}  | \psi_{E} \rangle & = &   \int_0^\infty dx \, \psi_{E_0}^*(x) \, \psi_{E}(x) = 0, \qquad E \in
{\cal C},  
\label{ap19}  \\ 
\langle  \psi_E | \psi_{E'} \rangle  & = &    \int_0^\infty dx \, \psi_E^*(x) \, \psi_{E'}(x) = 
\delta(E- E'), \qquad E, E'  \in {\cal C}.   
\label{ap20} 
\earray 
(\ref{ap18}) coincides with (\ref{ap16}). Eq. (\ref{ap19}), gives the relation between the coefficients
$A(E)$ and $B(E)$ of the wave function (\ref{ap11}), 

\beq
\frac{A(E)}{B(E)} = - \frac{ k_0^* - i k_+(E)}{  k_0^* - i k_-(E)}  = - 
\frac{ e^{ i  \vartheta}  - i \eta \,  e^{u}  }{e^{ i  \vartheta}  - i  \eta \,  e^{-u} },  
\label{ap21}
\eeq
where we have used eqs.(\ref{ap10}) and (\ref{ap14}). Condition (\ref{ap20}),  together with (\ref{ap21}),  fix the form
of these coefficients. Using  (\ref{ap11}) one obtains

\barray
& \langle  \psi_E | \psi_{E'} \rangle   =  A^*_E \, A_{E'} \left[ \pi \delta( k'_+ - k_+)   + i P \frac{ 1}{ k'_+ - k_+}  \right] + 
B^*_E \, B_{E'} \left[ \pi \delta( k'_- - k_-)   + i P \frac{ 1}{ k'_- - k_-}  \right]  & 
\nonumber 
 \\
 & +  A^*_E \, B_{E'} \left[ \pi \delta( k'_- - k_+)   + i P \frac{ 1}{ k'_-  - k_+}  \right] + 
B^*_E \, A_{E'} \left[ \pi \delta( k'_+  - k_-)   + i P \frac{ 1}{ k'_+  - k_-}  \right] &  \nonumber 
\earray 
where $k_\pm  = k_\pm(E), k'_\pm = k_\pm(E')$ and $P \frac{1}{x}$ denotes the principal part of $\frac{1}{x}$. 
To derive this equation we have used the improper  integral

\beq
\int_0^\infty dx \, e^{ i k x} = \pi \, \delta(k) + i P \frac{1}{k}, 
\label{ap23}
\eeq
which is the integral version of the distribution identity

\beq
\frac{1}{ k +  i 0} = -  i  \pi \delta(k) +  P \frac{1}{k}. 
\label{ap24}
\eeq
Eq.(\ref{ap20}) is satisfied provided

\barray
&  A^*_E \, A_{E'}  \delta( k_+ - k'_+)  + B^*_E \, B_{E'}  \delta( k_- - k'_-)  \label{ap25} \\
& +  
  A^*_E \, B_{E'}  \delta( k_+ - k'_-) + B^*_E \, A_{E'}  \delta( k_-  - k'_+) =
 \frac{1}{\pi} \delta(E- E'),  &
 \nonumber
 \earray
and

\beq
( A^*_E, B^*_E) \; M  \; \left( 
\begin{array}{c}
A_{E'} \\
B_{E'}
\end{array}
\right) = 0,
\label{ap26}
\eeq
where $M$ is the matrix 

\beq
  M = 
\left( 
\begin{array}{cc} 
\frac{1}{k_+ - k'_+} & \frac{1}{k_+ - k'_-} \\ 
\frac{1}{k_- - k'_+} & \frac{1}{k_- - k'_-} \\
\end{array} 
\right)  = \frac{\hbar}{ \ell_p} 
\left( 
\begin{array}{cc} 
\frac{1}{\eta \, e^{u}  - \eta'  e^{u'} } & \frac{1}{\eta \, e^{u}  - \eta'  e^{-u'} } \\ 
\frac{1}{\eta \, e^{-u}  - \eta'  e^{u'} } & \frac{1}{\eta \, e^{-u}  - \eta'  e^{-u'} } \\ 
\end{array} 
\right). 
\label{ap27}
\eeq
Using

\beq
\delta(k_\pm  - k'_\pm) =
 \frac{ \sqrt{E^2 - 4 \ell_p^2}}{ |k_\pm|} \delta(E - E'), \qquad 
 \delta(k_\pm  - k'_\mp) = 0, 
\label{ap28}
\nonumber 
\eeq
eq (\ref{ap25}) becomes 

\beq
|A_E|^2 \, \frac{ \sqrt{E^2 - 4 \ell_p^2}}{ |k_+|} + |B_E|^2  \, \frac{ \sqrt{E^2 - 4 \ell_p^2}}{ |k_-|} = \frac{1}{\pi}, 
\label{ap29}
\eeq
and similarly

\beq
|A_E|^2 \, e^{-u} + |B_E|^2 \, e^u = \frac{1}{ 2 \pi \hbar}. 
\label{ap30}
\eeq
Finally,  eq. (\ref{ap21}) implies

\barray
A(E) & = & \frac{ e^{ i \vartheta} - i  \eta \,  e^{u} }{ \sqrt{ 8 \pi \hbar ( \cosh u - \eta \, \sin \vartheta)} }, \label{ap31} \\
B(E) & = &  - \frac{ e^{ i \vartheta}  -  i  \eta \,  e^{-u} }{ \sqrt{ 8 \pi \hbar ( \cosh u - \eta \, \sin \vartheta)} }. 
\nonumber 
\earray 

These expressions satisfy eq.(\ref{ap26}), which finally proves  that the wave functions (\ref{ap11}), with coefficients
given by (\ref{ap31}), together with the normalizable state  (\ref{ap12}) form an orthonormal basis. 
The operator  $\hat{H}$,  has the physical meaning of momentum rather than energy. Its spectrum (\ref{ap8})
almost coincide with  that of the momentum operator $\hat{p}$ defined on the entire line, except in an interval around the origin $(- 2 \ell_p, 2 \ell_p)$,
which is replaced  by a bound state localized at the edge of the system. This result
is reminiscent of the Kondo model where a bound state is formed between an impurity
localized at  the origin and the conduction electrons \cite{GBT04}. However  in our model there is no spin
and it does not describe  a many body system,  so this analogy is for the time being formal.

\end{document}